\documentclass[journal]{IEEEtran}

\usepackage{cite}
\usepackage{amsmath,amsfonts,amsthm,amssymb}
\usepackage{graphicx}
\usepackage[usenames,dvipsnames]{xcolor} 
\usepackage{algorithm}
\usepackage[noend]{algpseudocode}
\usepackage{array}
\usepackage{url}
\usepackage{bm}
\usepackage{hyperref}
\usepackage{yfonts}
\usepackage{tabularx} 
\usepackage{tikz}
\usepackage{framed,enumitem}
\usepackage[OT1]{fontenc} 
\usetikzlibrary{matrix,decorations.pathreplacing,calc, positioning, fit}
\makeatletter
\let\MYcaption\@makecaption
\makeatother
\usepackage[font=footnotesize]{subcaption}
\makeatletter
\let\@makecaption\MYcaption
\makeatother

\newcommand{\C}{\mathcal{C}}
\newcommand{\Ce}{\mathtt{C}}
\newcommand{\N}{\mathcal{N}}
\newcommand{\A}{\mathtt{A}}
\newcommand{\mA}{\mathtt{A}}
\newcommand{\mB}{\mathtt{B}}
\newcommand{\I}{\mathcal{I}}
\newcommand{\B}{\mathcal{B}}
\newcommand{\D}{\mathcal{D}}
\newcommand{\E}{\mathcal{E}}
\newcommand{\K}{\mathcal{K}}

\newcommand{\X}{\mathtt{X}}
\newcommand{\Z}{\mathcal{Z}}
\renewcommand{\S}{\mathcal{S}}
\newcommand{\x}{\bm{x}}
\newcommand{\y}{\bm{y}}
\newcommand{\z}{\bm{z}}

\newcommand{\f}{{f}}
\renewcommand{\k}{\mathtt{k}}
\renewcommand{\j}{\mathtt{j}}

\newcommand{\p}{\mathtt{p}}
\newcommand{\q}{\mathtt{q}}
\newcommand{\Bst}{\textcolor{blue}{$\mathtt{B}$}}
\newcommand{\Yst}{\textcolor{orange}{$\mathtt{Y}$}} 
\newcommand{\Rst}{\textcolor{red}{$\mathtt{R}$}}
\newcommand{\1}{\mathtt{1}}
\newcommand{\2}{\mathtt{2}}

\newcommand{\R}{\mathfrak{R}}
\newcommand{\F}{\mathfrak{F}}
\newcommand{\Fe}{\mathcal{F}}
\newcommand{\argmax}{\operatornamewithlimits{argmax}}
\newcommand{\argmin}{\operatornamewithlimits{argmin}}

\newtheorem{theorem}{Theorem}
\newtheorem{definition}{Definition}
\newtheorem{example}{Example}
\newtheorem{lemma}{Lemma}
\newtheorem{remark}{Remark}
\newtheorem{assumption}{Assumption}
\newtheorem{proposition}{Proposition}
\newtheorem{corollary}{Corollary}
\newtheorem{problem}{Problem}
\usepackage{mathtools}
\DeclarePairedDelimiter{\ceil}{\lceil}{\rceil}

\begin{document}
\title{Equilibration Analysis and Control of Coordinating Decision-Making Populations}

\author{Negar Sakhaei, Zeinab Maleki and Pouria Ramazi%
\thanks{
This paper is accepted in part for presentation at, and publication in the proceedings of, the 60th IEEE Conference on Decision and Control, Austin, Texas, USA, December 13-15, 2021 \cite{Sakhaei2021equilibration}.
This paper additionally includes six numerical simulations, elaborations and figures on examples, Subsection \ref{sec:ri} on the rational imitation update rule, Section \ref{sec_supermodularity} on supermodularity, Section \ref{sec:publicGood} on public goods games, Subsections \ref{sec:cui} and \ref{sec:urc} on convergence under incentive and uniform reward control, the Appendix including Lemma \ref{proofOfExample:coo}, and the proofs of the results. 
N. Sakhaei and  Z. Maleki are with
the Department of Electrical and Computer Engineering, Isfahan University of Technology, Isfahan, Iran,
{\tt\small negar.sakhaei@yahoo.com, zmaleki@iut.ac.ir,}
P. Ramazi is with
the Department of Mathematics and Statistics, Brock University, Canada,
{\tt\small pramazi@brocku.ca}.}%
}
\maketitle

\begin{abstract}
    Whether a population of decision-making individuals will reach a state of satisfactory decisions is a fundamental problem in studying collective behaviors. 
    In the framework of evolutionary game theory and by means of potential functions, researchers have established equilibrium convergence under different update rules, including best-response and imitation, by imposing certain conditions on agents' utility functions. 
    Then by using the proposed potential functions, they have been able to control these populations towards some desired equilibrium. 
    Nevertheless, finding a potential function is often daunting, if not near impossible. 
    We introduce the so-called \emph{coordinating agent} who tends to switch to a decision only if at least another agent has done so. 
    We prove that any population of coordinating agents, a \emph{coordinating population}, almost surely equilibrates. 
    Apparently, some binary network games that were proven to equilibrate using potential functions are coordinating, and some coloring problems can be solved using this notion. 
    We additionally show that any mixed network of agents following  best-response, imitation, or rational imitation, and associated with coordination payoff matrices is coordinating, and hence, equilibrates. 
    As a second contribution, we provide an incentive-based control algorithm that leads coordinating populations to a desired equilibrium. 
    The algorithm iteratively maximizes the ratio of the number of agents choosing the desired decision to the provided incentive. It performs near optimal and as well as specialized algorithms proposed for best-response and imitation; however, it does not require a potential function. 
    Therefore, this control algorithm can be readily applied in general situations where no potential function is yet found for a given decision-making population. 
\end{abstract}
\begin{IEEEkeywords}
Decision-making populations, decision-making dynamics, coordinating, network games,  best-response, imitation, equilibration, convergence, control.
\end{IEEEkeywords}
\section{Introduction} 
Decision-making populations are evident in real-life, from choosing smartphone apps \cite{dogruel2015choosing} and following non-pharmaceutical policies \cite{yang2021effect}, to cooperative interactions between organisms \cite{coor} and exchanges within a cancer tumor \cite{cancer}. 
A natural question is whether these populations eventually reach an \textit{equilibrium}, where all individuals are satisfied with their choices.
A follow-up question is whether it is possible to ``shift'' the population from a state, possibly an equilibrium, to a more desired one. 
For example, being able to control the exchanges of diffusible goods within a cancer tumor may lead to the design of evolution-proof therapies that destroy the tumor \cite{cancer}. 
Similarly, traffic jams may be reduced by providing incentives to drivers to avoid peak hours \cite{kumar2016impacts}, and a disease spread may be mitigated by non-pharmaceutical interventions \cite{control:epidemic}.

A decision-making population can be modeled by a group of interacting agents who choose between some available options according to some revision protocol over time.
The evolution of the agents' choices results in the decision-making dynamics. 
Should the agents be linked via a \emph{network} and play two-player normal games with their neighbors, the population becomes a \emph{network game} \cite{survey:netgames, networkEGT}.
The two main revision protocols that have been considered previously are \textit{best-response} \cite{bopardikar2017convergence,matsui1992best} and \textit{imitation} \cite{barreiro2018constrained,tan2016analysis, fudenberg2008monotone}, the convergence analysis and control of which have been studied in various settings \cite{ramazi2016networks, granovetter1978threshold, ramazi2020convergence, durand2020controlling, durand2020optimal}.

To prove equilibrium convergence, the convention in systems and control is to find a potential (Lyapunov) function \cite{pot} and show that the equilibrium is asymptotically stable. 
The notion of \emph{potential game} is correspondingly defined and used in the literature \cite{monderer1996potential, marden2009joint, hao2018cooperative, cheng2014finite, cheng2018boolean, marden2012state, swenson2018best}. 
The potential function is also used to design a controller that leads the population to a desired equilibrium \cite{ramazi2020convergence, ramazi2016networks, li2019control}, for example, by changing the strategy of some agents \cite{optcontrol}, offering a uniform amount of incentive to all agents, and targeting specific agents and rewarding them \cite{control:bestresponse, control:imitation}.
However, finding this potential function is often overly challenging.

In this context, the notion of $\mathtt{A}$-coordinating network games was introduced in \cite{control:imitation} as a 2-strategy network in which if more agents play $\mathtt{A}$, all agents who tended to choose $\mathtt{A}$ will still do so.
It was proven that in an $\mathtt{A}$-coordinating network game that is at equilibrium, after offering payoff incentives for playing $\mathtt{A}$, no agent will switch away from $\mathtt{A}$ and the network reaches a unique new equilibrium -- the uniquely convergent property. 
Whether $\mathtt{A}$-coordinating networks equilibrate when starting from an arbitrary initial condition, remains an open problem though. 

We introduce a class of decision-making populations, a \textit{coordinating population}, consisting of agents who tend to pick a new choice only if some who were not picking that choice now do.
We show that this definition encompasses the previous, more restrictive definition of $\mathtt{A}$-coordinating populations.
We then establish that all coordinating populations almost surely reach an equilibrium state if the agents become active in an arbitrary order.    
Interestingly, a coordinating population may not converge under any activation sequence, implying that they do not admit a potential function, drawing a clear line from classical potential-function based analysis in control. 
Then, we consider network games as a special case of decision-making populations, provide conditions for several update rules to make a network game coordinating. 
For the first time, we demonstrate that any mixed network of imitators, best-responders, and rational imitators, associated with coordination payoff matrices, is coordinating, and therefore, almost surely equilibrates.
Moreover, similar to $\mathtt{A}$-coordinating populations, coordinating populations have the uniquely convergent property.
Then we design the \emph{iterative number to reward optimization (INRO)} control algorithm that approximates the amount of incentive needed to be provided to the agents in order to drive the population to a desired equilibrium.
The algorithm is not based on a potential function and can be applied to any decision-making dynamics, although convergence is guaranteed only for coordinating populations.
We show how this potential-free algorithm performs close to, and sometimes better than, existing other algorithms, including the IPRO algorithm that was proposed using potential functions on the best-response dynamics \cite{control:bestresponse}.

\section{Coordinating populations}
We consider a population of decision-making agents $\N = \{1, \dots, n\}$ who choose from the finite set of available decisions (choices) $\C$ over time $t\in\mathbb{Z}_{\geq0}$.
We stack all agents' choices in the \textit{(decision) state} $\bm{x} = (x_1, \dots, {x}_n)^\top$, where $x_i\in\C$ denotes the choice of agent $i$.
Each agent is initially associated with a choice.
At every time $t\geq0$, an agent $i$ becomes active to revise her choice based on the decision state $\x(t)$ and according to the \textit{revision protocol} $\F = (\f_1, \dots, \f_n)^\top, \f_j: \C^n \rightarrow \C$, where $\f_j(\x)$ is what agent $j$ tends to choose. 
More specifically, agent $i$ active at time $t$ revises her choice at time $t+1$ to
\begin{equation}   \label{eq:updatedstrategy}
     x_i(t+1) = \f_i(\x(t)),
\end{equation}
and the remaining agents do not change their choices at time $t+1$.
The triple $(\N, \C, \F)$ defines a \textit{(decision-making) population}.

The agents become active according to an \textit{activation sequence} $\mathcal{A} = (a_t)_{t=0}^{\infty}$, where $a_t \in \N$ is the active agent at time $t$.
The sequence is \textit{asynchronous} as only one agent is active at a time.
The activation sequence can be deterministic or randomly generated and even depend on the agents' previous choices. 
The revision protocol $\F$ and activation sequence $\mathcal{A}$ govern the evolution of the decision state $\x(t)$, which we refer to as the \textit{decision-making dynamics}.

\begin{figure}[ht]
    \centering
    \includegraphics[width=\linewidth]{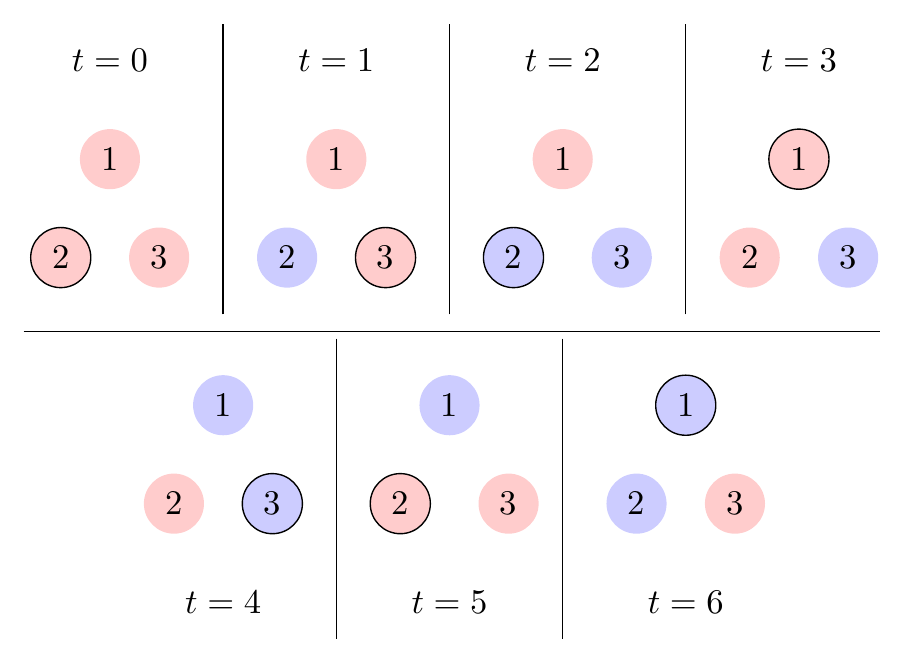}
    \caption{%
        \textbf{The evolution of the decision state under the non-equilibrating decision-making dynamics in Example \ref{example_nonEquilibration}.} 
        Circles $1$, $2$, and $3$ represent the corresponding agents, and colors $\mathtt{\textcolor{blue}{Blue}}$ and $\mathtt{\textcolor{red}{Red}}$ represent the corresponding choices.
        The active agent is indicated by a solid outline.
        Under the activation subsequence $(2, 3, 2, 1, 3, 2, 1)$, 
        the population visits seven of the eight total possible states and at time $t=7$ returns to its state at time $t=1$.
        The state perpetually switches between these six states and never equilibrates.%
    }
    \label{fig:noteq}
\end{figure}

\begin{figure}[!ht]
    \centering
    \begin{subfigure}[b]{0.45\textwidth}
        \includegraphics[width=\linewidth]{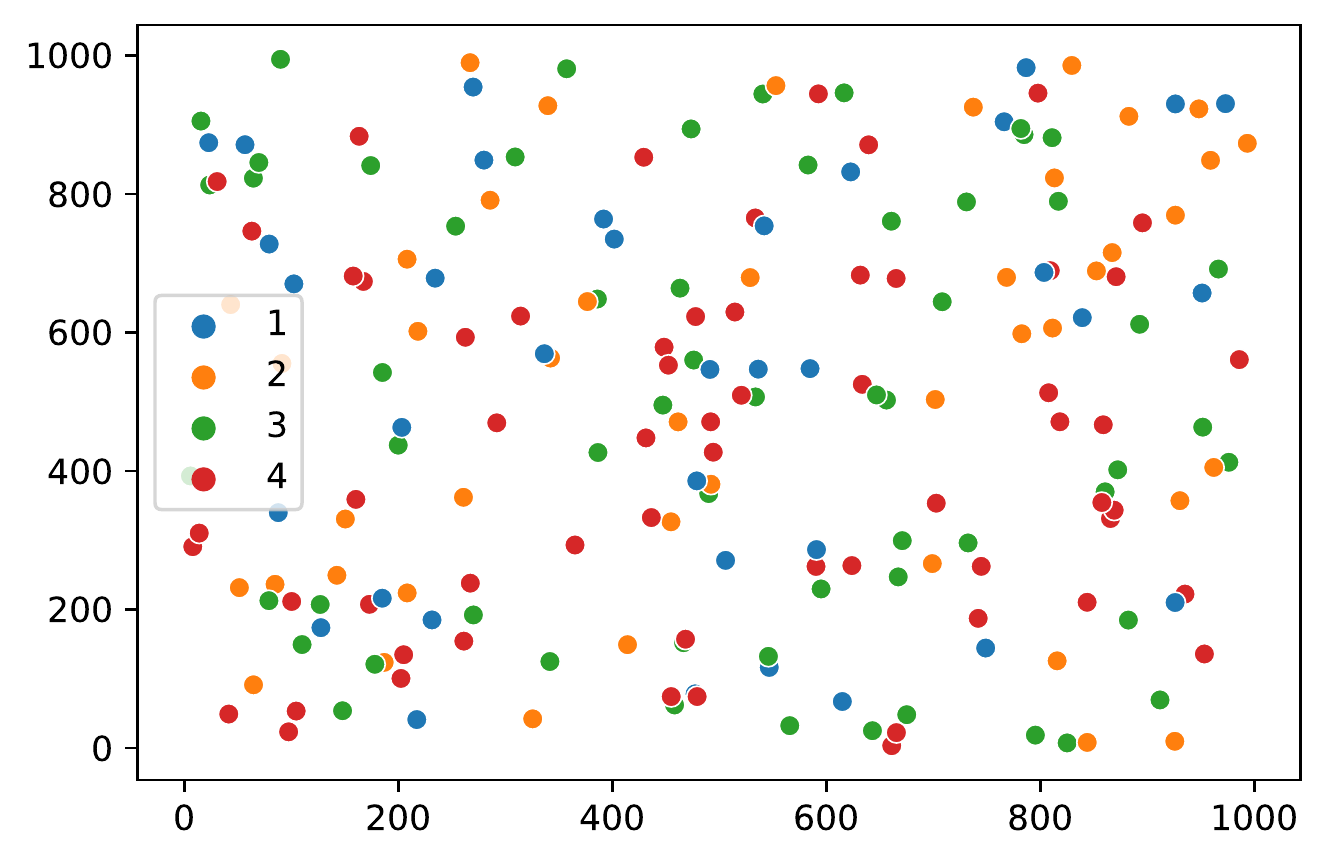}
        \caption{The population at state $\bm{x}(0)$.}
    \end{subfigure}
    \begin{subfigure}[b]{0.45\textwidth}
            \includegraphics[width=\linewidth]{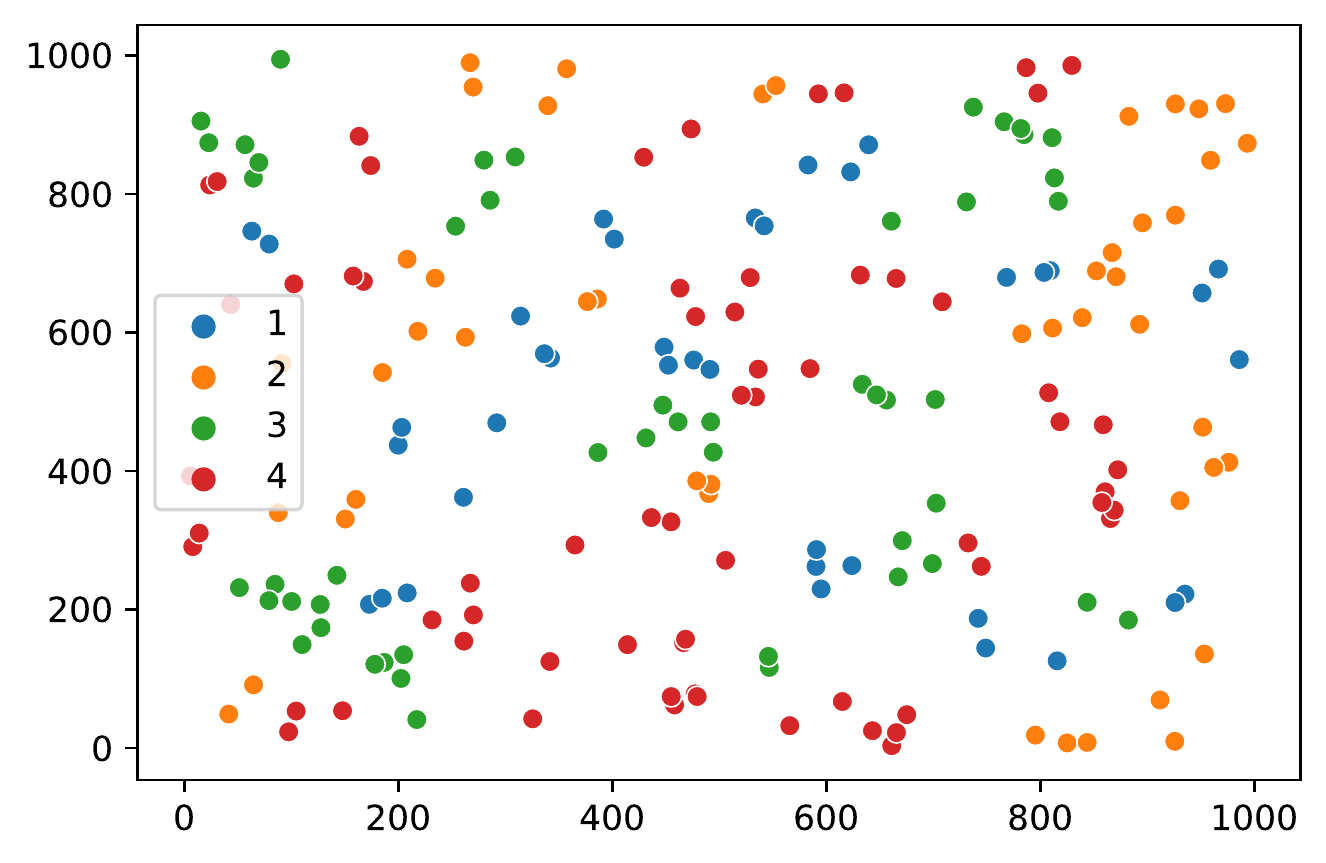}
        \caption{The population at state $\bm{x}^*$.}
    \end{subfigure}
    \begin{subfigure}[b]{0.45\textwidth}
            \includegraphics[width=\linewidth]{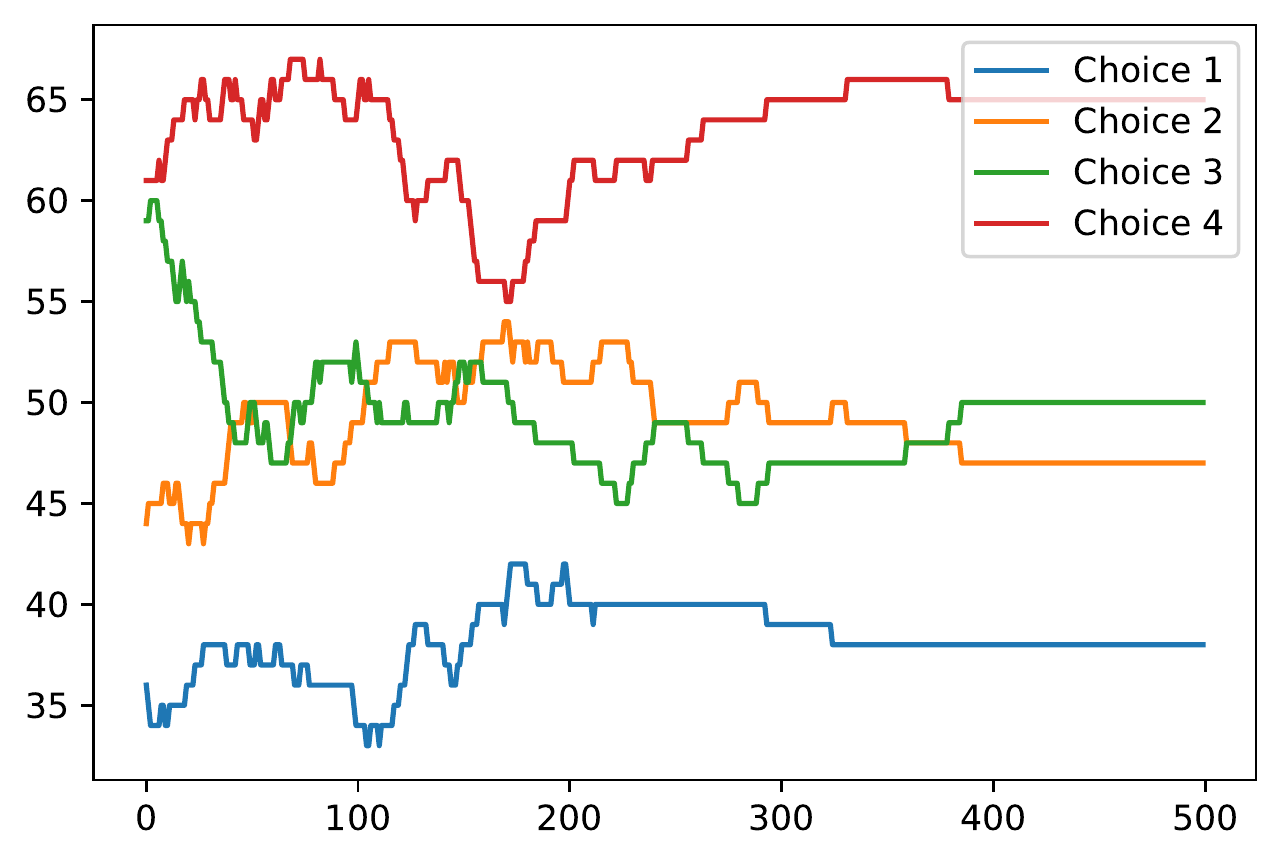}
        \caption{The evolution of the number of agents choosing each option over time.}
    \end{subfigure}
    \caption{\textbf{The evolution of the decision state under the equilibrating decision-making dynamics in Example \ref{example:coo}}. 
    The population consists of 200 agents, each choosing between four options. It starts from the state $\x(0)$, in which 36 agents choose $\textcolor{blue}{\1}$, 44 choose $\textcolor{orange}{\2}$, 59 choose $\textcolor{PineGreen}{\mathtt{3}}$ and 61 choose $\textcolor{red}{\mathtt{4}}$. 
    At every time step, one agent activates and switches her strategy to the one chosen by her nearest neighbor. 
    These updates continue until clusters of same-color agents appear in the plane, such that the nearest neighbor of each agent in the cluster is included in the cluster. 
    This results in the equilibrium state $\bm{x}^*$, in which 38 agents choose $\mathtt{\textcolor{blue}{1}}$, 47 choose $\mathtt{\textcolor{orange}{2}}$, 50 choose $\mathtt{\textcolor{PineGreen}{3}}$ and 65 choose $\mathtt{\textcolor{red}{4}}$.}
    \label{fig:eqex}
\end{figure}

Our goal is to investigate the equilibration of the decision-making dynamics -- given an initial condition $\x(0)$, whether the state $\x(t)$ eventually reaches, at some finite time $t^*$, an equilibrium state $\x^*$ at which no agent tends to change her choice, i.e., $\x(t) = \x^*$  for all $t \geq t^*$.
The decision-making dynamics do not equilibrate under every revision protocol, as illustrated in the following example. 
\begin{example} \label{example_nonEquilibration}
    Consider the population of agents $\N = \{1, 2, 3\}$ with available choices $\C = \{\mathtt{\textcolor{red}{Red}}, \mathtt{\textcolor{blue}{Blue}}\}$ and revision protocol $ \F = (\f_i)^n_{i=1}$, $\f_i(\x) = \argmin_{\k \in \C} n_{\k}(\x)$, where $n_{\k}(\x)$ is the number of agents that have choice $\k$ at state $\x$.
    So the agents update to the least common color.
    This population will never equilibrate, because starting from any initial state and under any activation sequence, after at most one time step the state will keep alternating between two outcomes: one where a single agent has chosen $\mathtt{\textcolor{blue}{Blue}}$ and the other where a single agent has chosen $\mathtt{\textcolor{red}{Red}}$ (\autoref{fig:noteq}). 
\end{example}
However, some decision-making populations equilibrate.
\begin{example} \label{example:coo}
    Consider the population defined by $(\N, \{\textcolor{blue}{\1}, \textcolor{orange}{\2}, \textcolor{PineGreen}{\mathtt{3}}, \textcolor{red}{\mathtt{4}}\}, \F)$, where each agent $i \in \N$ represents a point $p_i$ in the $2D$ plane, and
    $$\f_i(\x) = x_j, \qquad j = \argmin_{k \in \N} d(p_i,p_k),$$
    where $d(x,y)$ is the Euclidean distance between the points $\bm{x}$ and $\bm{y}$.
    That is, the agents update to the choice of their nearest neighbor (Figure \ref{fig:eqex}).
    We assume that every pair of agents have a unique distance from each other.
    This population can be shown to reach an equilibrium under any activation sequence and starting from any initial condition (Lemma \ref{proofOfExample:coo} in the appendix).  
    The equilibrium state is not unique though. 
    For example, every state where all agents have the same choice is an equilibrium.
\end{example}
Intuitively, the revision protocol in Example \ref{example:coo} has the property that the agents are ``more likely'' to choose $\A$ if some of those who were not selecting $\A$, have now chosen $\A$.
In what follows, we rigorously define and extend this property to a class of revision protocols, which we later show to guarantee equilibration of the decision-making dynamics. 
Given the states $\bm{y}, \bm{z} \in \C^n$, define $\K(\z,\y)\subseteq\C$ as the set of every choice $\A$ for which there exists an agent that selects $\A$ at $\z$ but does not select $\A$ at $\y$, i.e., 
\begin{equation*}    \label{coordinatingDefinitionEq-1}
    \K(\z,\y) \triangleq 
    \left\{\A\in\C\,|\, \exists i\in\N: (z_i = \A \wedge  y_i \neq\A)\right\}.
\end{equation*}
\begin{definition}[Coordinating population] \label{acoodef}
    Given the population defined by $(\N, \C, \F)$, agent $i\in\N$ is \emph{coordinating} if for any two decision states $\bm{y}, \bm{z} \in \C^n$, 
    \begin{equation}    \label{coordinatingDefinitionEq-2}
        \f_i(\z) \in \{\f_i(\y)\}\cup\K(\z,\y).%
    \end{equation}
    The corresponding revision protocol $\f_i$ is called a \emph{coordinating revision protocol},
    and a population of all coordinating agents is called a \emph{coordinating population}. 
\end{definition}
Given state $\y$, the choice of a coordinating agent $i$ at another state $\z$ may be either her choice at $\y$ or any of the choices that an agent has exclusively selected at $\z$.
So a coordinating agent switches her tendency towards a choice only if some agent(s) have switched to that choice.
Note that a coordinating population does not ensure monotonic increments of the number of $\A$-selectors for a given $\A\in\C$.
It only ensures that an increase in the number of agents choosing $\A$ will never cause an $\A$-selecting agent to switch to another choice, and may cause other agents (not choosing $\A$) to switch to $\A$.
The coordinating property turns out to be common in many known decision-making populations, e.g., network games, as we explore in Sections \ref{sec:netgames} and \ref{sec:publicGood}. 
\subsection{Coordinating versus $\mathtt{A}$-coordinating protocols}
Our definition of a coordinating population is inspired by that of \emph{$\A$-coordinating} network games \cite{riehl2017control}, also known as \emph{coordinating in $\X$} network games \cite{riehl2018survey}, defined in the context of evolutionary game theory. 
We rephrase the definition to match our general setup of decision-making populations.
\begin{definition}[$\mathtt{A}$-coordinating population \cite{riehl2017control,riehl2018survey}] \label{acoodef2}
    Consider the population defined by $(\N, \C, \F)$. 
    Agent $i\in\N$ is \emph{$\mathtt{A}$-coordinating}, $\mathtt{A} \in \C$, if for any two decision states $\bm{y}, \bm{z} \in \C^n$ satisfying
    \begin{equation}    \label{ACoordinatingDefinitionEq1}
        y_j = \mathtt{A} \Rightarrow z_j = \mathtt{A} \quad \forall j \in \N,
    \end{equation}
    the following holds:
    \begin{equation}
        \f_i(\bm{y}) = \mathtt{A} \Rightarrow \f_i(\bm{z}) = \mathtt{A},  \label{ACoordinatingDefinitionEq2} 
    \end{equation}   
    A population of all $\mathtt{A}$-coordinating agents is called an $\mathtt{A}$-coordinating population.
\end{definition}
States $\y$ and $\z$ are such that all agents that choose $\A$ at $\y$ also do so at $\z$.
Condition \eqref{ACoordinatingDefinitionEq2} states that if an agent tends to choose $\mathtt{A}$ at $\bm{y}$, then it has to choose $\mathtt{A}$ also at $\bm{z}$.
The definition is for a particular choice $\mathtt{A}$, but can be extended to what we call a \emph{restrictive coordinating agent} which is an agent that is $\mathtt{A}$-coordinating for every choice $\mathtt{A}\in\C$.
Correspondingly, a \emph{restrictive coordinating population} is a population of all restrictive coordinating agents.
The restrictiveness can be explained as follows. 
According to \eqref{ACoordinatingDefinitionEq1} and \eqref{ACoordinatingDefinitionEq2}, if all who have chosen $\mathtt{A}$ at $\y$ do the same at $\z$, then any agent who tends to choose exclusively $\mathtt{A}$ at $\y$ will do the same at $\z$.
However, for a coordinating population, \eqref{coordinatingDefinitionEq-2} allows the agents to additionally choose other choices in $\K(\z,\y)$.
Although the definitions of the two populations match for binary choices, i.e., $|\C|=2$, coordinating populations are more general.
\begin{proposition} \label{restrictiveCoordinatingIsCoordinating}
    A restrictive coordinating population is a coordinating population.
\end{proposition}
\begin{IEEEproof}
    We prove by contradiction.
    Consider a restrictive coordinating population defined by $(\N, \C, \R)$.
    Assume on the contrary, the existence of two decision states $\y,\z\in\C^n$ that do not satisfy  \eqref{coordinatingDefinitionEq-2} for some agent $i\in\N$; that is,
    \begin{equation}    \label{comparisonOfDefinitions}
        f_i(\z)\neq f_i(\y) \wedge f_i(\z)\not\in\K(\z,\y). 
    \end{equation}
    Let $f_i(\z) = \mathtt{A}$. 
    The exclusion $f_i(\z)\not\in\K(\z,\y)$ implies that if an agent has chosen $\mathtt{A}$ at $\z$, she has also done so at $\y$:
    \begin{equation*}
        z_j = \mathtt{A}
        \Rightarrow y_j = \mathtt{A} \quad \forall j\in\N.
    \end{equation*}
    Thus, according to Definition \ref{acoodef2}, for every agent, including agent $i$, it holds that
    \begin{equation*}
        \f_i(\z) = \mathtt{A}
        \Rightarrow \f_i(\y) = \mathtt{A}.
    \end{equation*}
    Hence, $\f_i(\y) = \mathtt{A}$, yielding $\f_i(\z) = \f_i(\y)$, contracting \eqref{comparisonOfDefinitions}, completing the proof. 
\end{IEEEproof}

\begin{figure}[!ht]
    \centering
    \begin{subfigure}[b]{0.45\textwidth}
        \includegraphics[width=\linewidth]{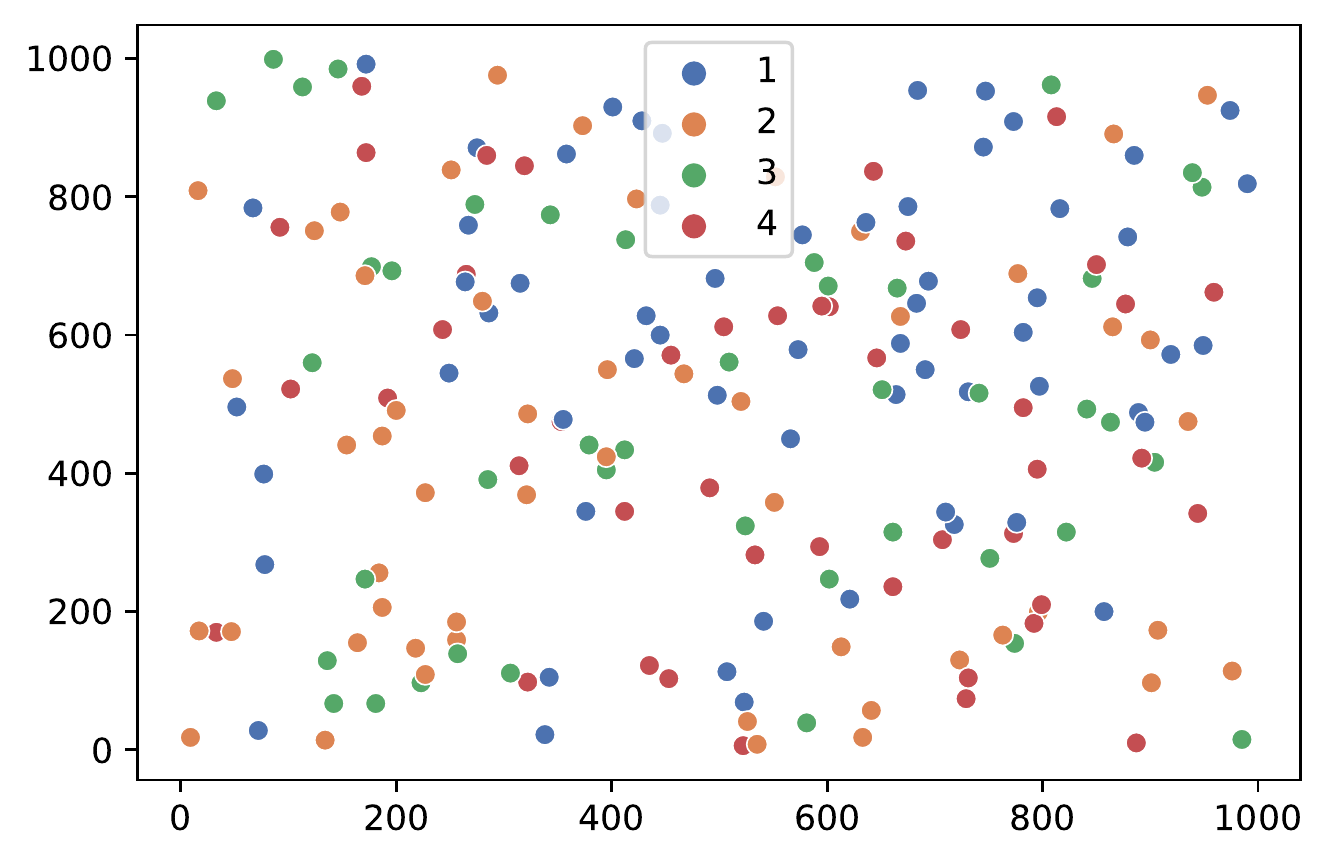}
        \caption{The population at state $\bm{x}(0)$.}
    \end{subfigure}
    \begin{subfigure}[b]{0.45\textwidth}
            \includegraphics[width=\linewidth]{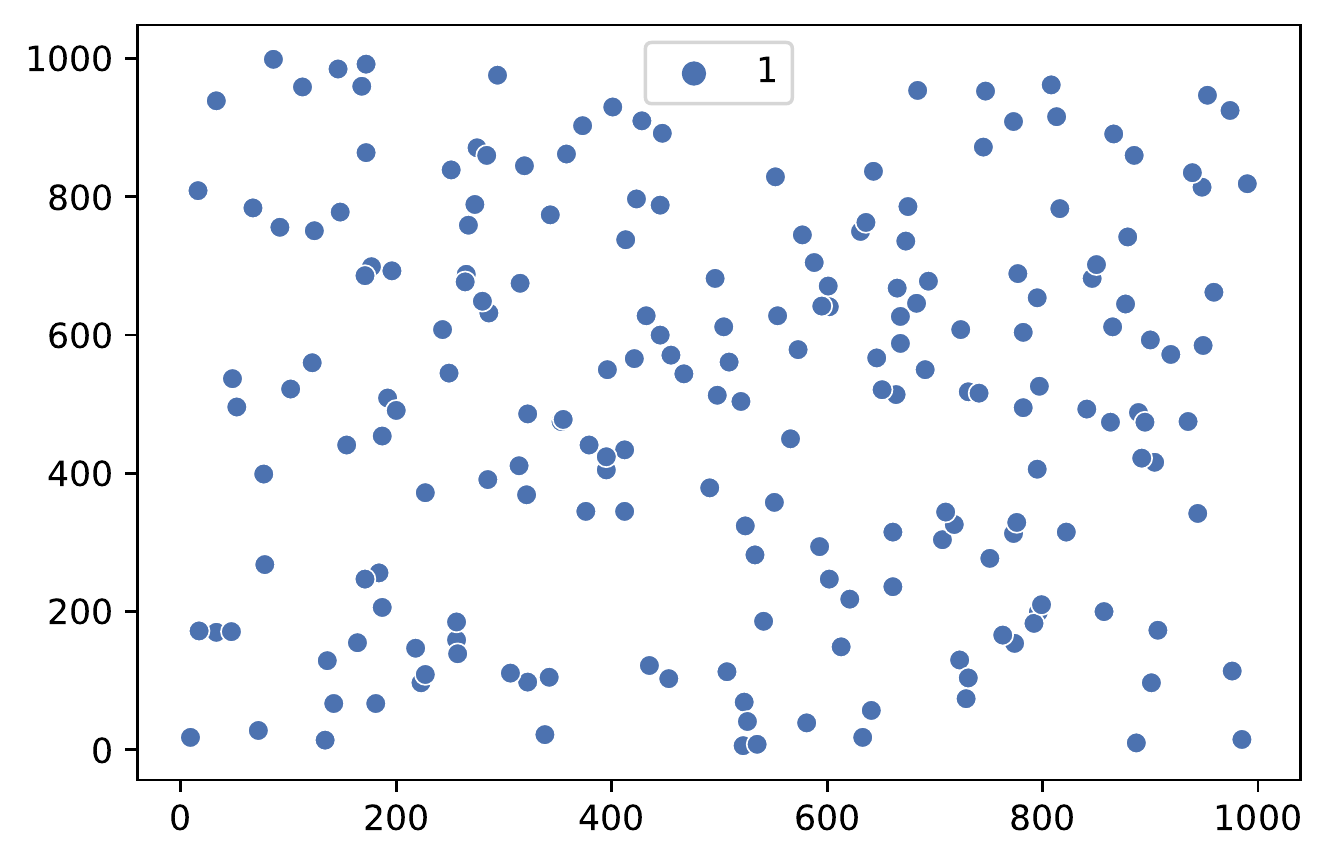}
        \caption{The population at state $\bm{x}^*$.}
    \end{subfigure}
    \begin{subfigure}[b]{0.45\textwidth}
            \includegraphics[width=\linewidth]{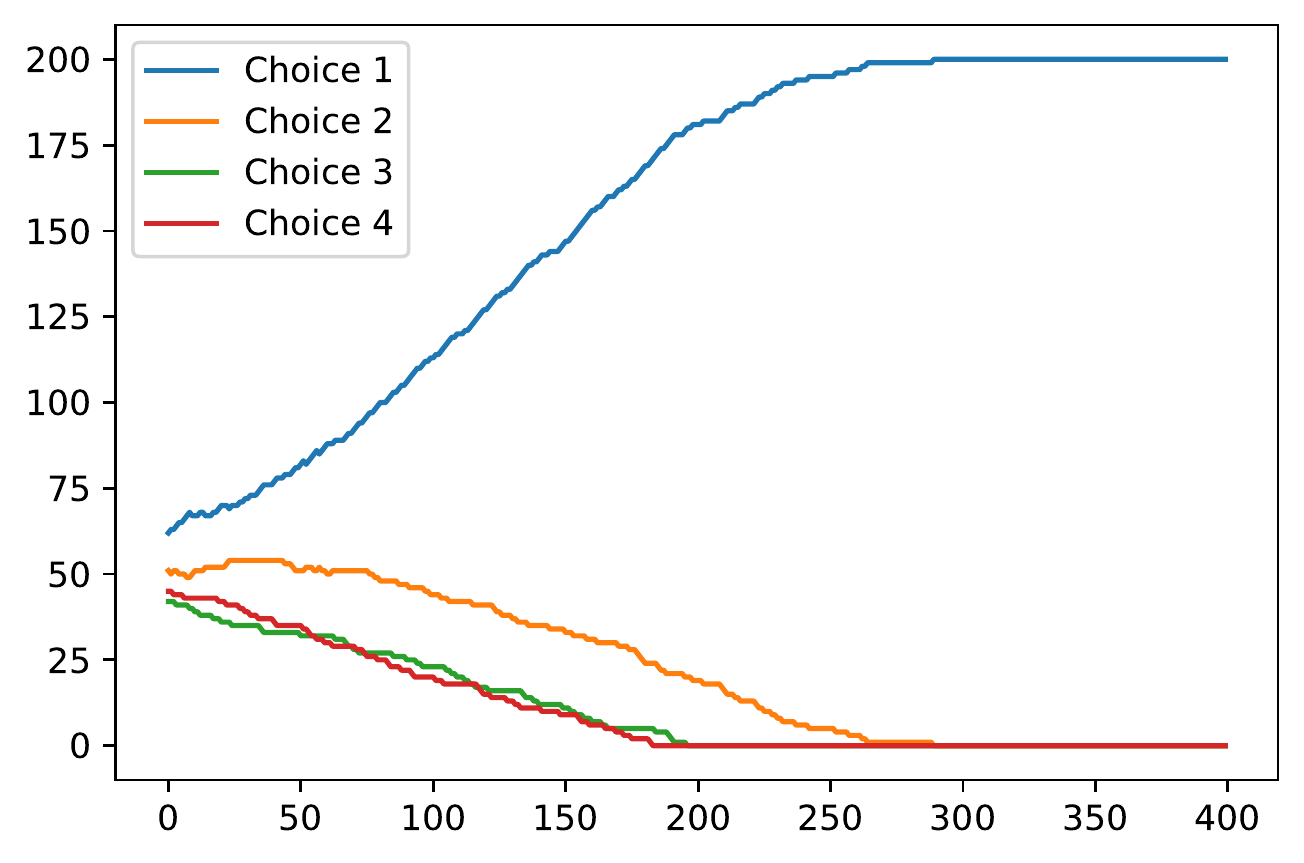}
        \caption{The evolution of the number of agents choosing each option over time.}
    \end{subfigure}
    \caption{\textbf{The evolution of the decision state under the equilibrating decision-making dynamics in Example \ref{example:coo}}. 
    The population consists of 200 agents, each choosing between four options. It starts from the state $\x(0)$, in which 66 agents choose $\textcolor{blue}{\1}$, 51 choose $\textcolor{orange}{\2}$, 40 choose $\textcolor{PineGreen}{\mathtt{3}}$ and 43 choose $\textcolor{red}{\mathtt{4}}$. 
    At every time step, one agent activates and switches her strategy to the one chosen most frequently by all her neighbors that are located within a distance of 50. 
    This results in the equilibrium state $\bm{x}^*$, in which all 200 agents choose $\mathtt{\textcolor{blue}{1}}$.}
    \label{fig:eqex2}
\end{figure}

The population in Example \ref{example:coo} is restrictive coordinating and hence also coordinating. 
The following modified version is an example of a coordinating population that is not restrictive. 
\begin{example} \label{example:coo_nonrestrictive}
    Consider the population in Example \ref{example:coo}, where the revision protocol is replaced with
    $$\f_i(\x) = \argmax_{j \in \N_i^{50}} n_{x_j}(\x).$$
    where $\N_i^{50}$ is the set of agents that fall into the disc of radius 50 centered at $p_i$.
    That is, the agents update to the most frequent choice of their neighbors that are located within a distance of 50 (Figure \ref{fig:eqex2}).
    This population equilibrates under any activation sequence and starting from any initial condition.  
    However, the population is not restrictive coordinating.
    For example, if within the aforementioned disc centered around agent $i$, choice $\textcolor{blue}{\1}$ is the most frequent, then changing the choices of enough agents in the disc from $\textcolor{orange}{\2}$ to $\textcolor{PineGreen}{\mathtt{3}}$ causes agent $i$ to tend to switch to $\textcolor{PineGreen}{\mathtt{3}}$ rather than $\textcolor{blue}{\1}$.
    This is despite the fact that no agent has changed her choice from $\textcolor{blue}{\1}$. 
\end{example}

For control purposes, it was shown in \cite{riehl2017control,control:bestresponse} that if an $\mathtt{A}$-coordinating population is already at equilibrium and some agents are provided incentives to switch their choices to $\mathtt{A}$, then the population will again equilibriate. 
However, no results are known for whether the population equilibrates if it starts from an arbitrary initial condition.
In Section \ref{sec_Convergence}, we investigate this question for a general coordinating population.

\section{Convergence}   \label{sec_Convergence}
In what follows, we construct a finite activation sequence that is independent from the initial state and drives the decision-making population to an equilibrium.
Let $\D_i=(i,i-1,\ldots,1), i\in\N$.
Define the activation sequence $\S$ by
\begin{equation*}
    \S \triangleq (\S_1,\S_2, \ldots, \S_n),
\end{equation*}
where
\begin{equation*}
    \begin{cases}
        \S_i = (i,\underbrace{\D_{i-1},\ldots,\D_{i-1}}_{i-1}), &i=2,\ldots,n\\
        \S_1 = (1)  &
    \end{cases}.
\end{equation*}
For example, for $n=3$, we obtain $\S = (1,2,1,3,2,1,2,1)$.
\begin{lemma} \label{lem_ACoordinatingEquilibrates}
    Every coordinating population equilibrates under $\S$.
\end{lemma}
\begin{IEEEproof}
    We prove by induction on $j\in\N$ that agents $1,2,\ldots,j$ are satisfied with their choices at time $T=\sum_{i=1}^j|\S_i|$.
    For $j=1$, note that agent $1$ will become active at time $t=0$ under $\S_1$. 
    If she does not switch, she is satisfied with her choice at time $T=|\S_1|=1$.
    So consider the case where she switches her choice from say $\mathtt{A}$ to $\mathtt{B}$ at $t=1$. 
    Then for the decision states $\y=\x(0)$ and $\z=\x(1)$, we have $f_1(\y) = \mathtt{B}$ and $\K(\z,\y) = \{\mathtt{B}\}$. 
    Therefore, since the population is coordinating, \eqref{coordinatingDefinitionEq-2} is in force, implying that
    $\f_1(\z) = \mathtt{B}$.
    Hence, agent $1$ does not tend to switch strategies at $\z=\x(1)$. 
    Thus, the statement holds for $j=1$.
    
    Assume that the inductive statement holds for $j = q\leq n-1$; that is, agents $1,\ldots,q$ do not tend to switch at time $t_1 = \sum_{i=1}^q|\S_i|$. 
    Under $S_{q+1}$, agent $q+1$ becomes active exactly once and at time $t_1+1$.
    If agent $q+1$ does not switch her choice at $t_1+1$, then agents $1,\ldots,q$ remain satisfied with their choices up to time $T = \sum_{i=1}^{q+1}|\S_i|$, because the decision state remains the same as it was at time $t_1$.
    So the inductive step is complete in this case.
    Otherwise, agent $q+1$ switches to some choice $\mathtt{B}\in\C$ at time $t_1+1$. 
    We prove by contradiction that then up to time $T$, agents may only switch to $\mathtt{B}$.
    Assume the contrary, and let $t_2\in\{t_1+2,\ldots,T\}$ be the first time when an agent $i\in\{1,\ldots,q\}$ switches to some other choice $\mathtt{A}\in\C$. 
    For the decision states $\y = \x(t_1)$ and $\z = \x(t_2-1)$, we have $\K(\z,\y) = \{\mathtt{B}\}$ and $f_i(\z) = \mathtt{A}$.
    On the other hand, since the population is coordinating, \eqref{coordinatingDefinitionEq-2} is in force, implying that
    $\mathtt{A} \in \{\f_i(\y)\} \cup \{\mathtt{B}\}$.
    Hence, $\f_i(\y) = \mathtt{A} = f_i(\z)$. 
    Moreover, $f_i(\y) = y_i$.
    Hence, agent $i$ tends to choose at time $t_2-1$, the same choice she had and was satisfied with at time $t_1$, meaning that she does not switch choices at time $t_2$, a contradiction.
    So up to time $T$, agents may only switch to $\mathtt{B}$.
    On the other hand, after time $t_1+1$, agents $q,q-1,\ldots,1$ will become active in order and $q$ times.
    So either at each time, at least one of the agents switches her choice to $\mathtt{B}$ or at some point all agents become and remain satisfied with their choices. 
    The first case can continue for at most $q$ switches, implying that agents $1,\ldots,q$ are again all satisfied by time $T$.
    So the inductive step is verified in both cases.
    Therefore, the statement holds for $j=n$, which completes the proof.
\end{IEEEproof}
Lemma \ref{lem_ACoordinatingEquilibrates} leads to equilibration under random activation sequences where the probability of any of the agents becoming active does not vanish over time, because then activation subsequence $\S$ is guaranteed to appear in the long run.
By a \textit{random activation sequence}, we mean a realization of the random variables $(A_t)_{t=0}^{\infty}$, where $A_t \in \N$ are mutually independent random variables with a positive support over $\N$.
We make the mild assumption that the support of each random variable $A_t$ is lower bounded by some positive constant. 
Let $P$ be the probability density function.
\begin{assumption}  \label{assumption1}
    There exists some constant $K>0$, such that for any agent $i\in\N$ and time $t\in\mathbb{Z}_{\geq0}$, $P(A_t = i)\geq K$.
\end{assumption}
This guarantees that any finite sequence of agents appears almost surely in the random activation sequence with probability one. 
The corresponding results hold ``almost surely''; that is, with probability one. 
Note that an event may happen almost surely even if it does not include all possible outcomes in the probability space. 
For example, a tail is almost surely flipped in tossing a coin repeatedly because the probability of having only heads becomes zero as the number of times the coin is flipped approaches infinity.  
\begin{theorem} \label{theorem_ACoordinatingEquilibrates}
    Every coordinating population equilibrates almost surely under any random activation sequence.
\end{theorem}
\begin{IEEEproof}
    The activation sequence $\S$, thus it appears in any random activation sequence with probability one as time approaches infinity.
    Upon appearance, the population will reach an equilibrium by the end of the sequence $\S$ in view of Lemma \ref{lem_ACoordinatingEquilibrates}.
    This completes the proof. 
\end{IEEEproof}
As an immediate result, both decision-making dynamics in Examples \ref{example:coo} and \ref{example:coo_nonrestrictive} almost surely equilibrate under a random activation sequence. 

\begin{figure}
    \centering
    \includegraphics{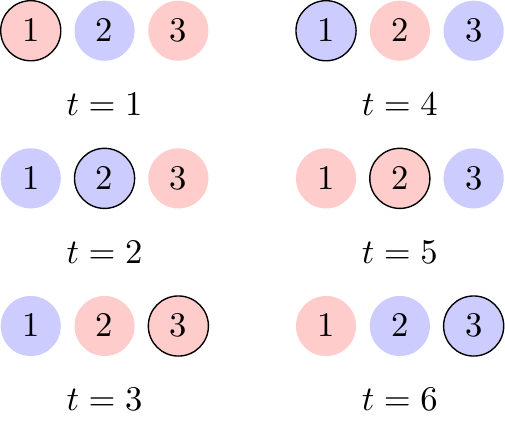}
    \caption{\textbf{The non-equilibrating evolution of the decision state under a coordinating update rule.} 
    The population ($\N=\{1, 2, 3\}, \F)$ with $\f_i = \{x_{i+1}\}$ is under the persistent activation sequence (1, 2, 3, 1, 2, 3, ...). The decision state loops through the indicated six states perpetually.}
    \label{fig:nonEqCoordinating}
\end{figure}

What about a non-random activation sequence that may never include $\S$?
The following is an example of a coordinating population that never equilibrates under a non-random activation sequence. 
\begin{example} \label{example_nonEquilibratingCoordinatingPopulation}
    Consider the population of $n\geq 3$ agents ($n$ being odd) with the set of choices $\C=\{\mA,\mB\}$ who follow the revision protocol $\F = (\f_i)_{i=1}^n, \f_i(\x) = x_{i+1}$ for $i\leq n-1$, and $\f_{n}(\x) = x_1$.
    Starting from the decision state $\x(0) = (\mA,\mB,\mA,\mB,\ldots,\mA)^\top$, and under the activation sequence $(\B,\B,\ldots)$, where $\B =(1,2,\ldots,n)$, the population never reaches an equilibrium. 
    See Figure~\ref{fig:nonEqCoordinating} for the case with $n=3$.
    
    Nevertheless, this population is coordinating:
    Given two arbitrary decision states $\y,\z\in\C^n$, for every agent $i$, either $z_{i+1} = y_{i+1}$ (module $n$) or $z_{i+1}$ is different from $y_{i+1}$, implying that agent $i+1$ has chosen a strategy at $\z$ that is different from her strategy at $\y$, yielding $z_{i+1}\in\K(\z,\y)$. 
    Hence, 
    \begin{equation*}
        z_{i+1} \in \{y_{i+1}\} \cup \K(\z,\y),
    \end{equation*}
    which results in \eqref{coordinatingDefinitionEq-2}.
    The population is also restrictive coordinating: 
    For any two decision states $\y,\z\in\C^n$ satisfying \eqref{ACoordinatingDefinitionEq1}, the following holds by letting $j=i+1$:
    \begin{equation*}
        z_{i+1}(\y) = \mathtt{A}
        \Rightarrow y_{i+1}(\z) = \mathtt{A},
    \end{equation*}
    which results in \eqref{ACoordinatingDefinitionEq2}, and the same holds for choice $\mathtt{B}$.
\end{example}
Therefore, a given coordinating population does not necessarily equilibrate under every activation sequence. 
Nevertheless, almost surely it does equilibrate under a random activation sequence, which is the case with most, if not all, realistic activation sequences.
\section{Examples in pairwise network games}\label{sec:netgames}
A particular case of decision-making populations is the so-called  \textit{(pairwise) network game} \cite{riehl2018survey,cheng2015modeling}, in which the agents are interconnected by a network $\mathbb{G} = (\N, \mathcal{E})$, with the edge set $\mathcal{E} \subseteq \N\times\N$.
The choice $x_i \in \C$ that an agent $i$ makes is the strategy she plays against each of her neighbors $\mathcal{N}_i \subset \N$ in two-player games. 
Upon playing with her neighbour $j \in \mathcal{N}_i$, agent $i$ earns a payoff $\pi^i_{x_i, x_j}$ where $\bm{\pi}^i \in \mathbb{R}^{|\C|\times |C|}$ is the \textit{payoff matrix} associated with agent $i$.
Consequently, the accumulated payoff or \textit{utility} of agent $i$ is
\begin{equation}\label{eq:totalpayoff}
    u_i(\x) \triangleq \sum_{j \in \mathcal{N}_i} \pi^i_{x_i, x_j}.
\end{equation}
At each time $t$, similar to the general framework, an agent becomes active to revise her strategy based on her own and/or her neighbor's utilities and according to some update rule $\R$. 
Unlike the revision protocol that provides the feasible action $f_i$ for active agent $i$, the update rule provides a feasible \textit{set} of actions $\Fe_i: \C^n \rightarrow 2^\C\backslash \emptyset$ from which the agent may choose her action.
The network game is then defined as the triple $\Gamma = (\mathbb{G}, \bm{\pi}, \R)$, where $\bm{\pi} = (\bm{\pi^1}, \dots, \bm{\pi^n})^\top$ is the stacking of the payoff matrices, and similarly, $\R = (\Fe_1,\ldots,\Fe_n)$.
We say a network game is \emph{binary}, if $|\C| = 2$.

Should there be more than one feasible choice, the active agent selects the one provided by the \textit{tie breaker} $\bm{\tau} = (\tau_1,\ldots,\tau_n)$, where $\tau_i:2^\C \backslash \emptyset\times \C^n\to \C$ puts forward a single choice for agent $i$ based on her feasible set $\Fe_i(\x)$ and the decision state $\x$, i.e., her feasible choice $f_i(\x)$ equals $\tau_i(\Fe_i(\x),\x)$. 
To ensure that the tie breaker belongs to the feasible set, we assume that $\tau_i(\mathcal{S},\cdot)\in \mathcal{S}$ for all nonempty $\mathcal{S}\subseteq \C$.
For example, if there is a preference over the choices, and they are labeled by $\1,\2,\ldots$, then $\tau_i(\Fe_i(\x),\x)$ can be the least of the feasible choices, i.e., $\min \Fe_i(\x)$; or in case of having binary choices, i.e., $|\C|=2$, and a tie, i.e., $|\Fe_i(\x)|=2$, then $\tau_i(\Fe_i(\x),\x)$ can be the agent's choice $x_i$ or that of some other agent $x_j$.
Note that $\tau_i(\mathcal{S},\cdot)$, $\mathcal{S}\subseteq\C$, matches the definition of a revision protocol. 
This naturally leads to the definition of a \emph{coordinating tie breaker} $\tau_i$ that for any two decision states $\bm{y}, \bm{z} \in \C^n$, and any $\S\subseteq\C$,
\begin{equation}    \label{coordinatingTieBreaker-1}
    \tau_i(\S,\z) \in \{\tau_i(\S,\y)\}\cup\K(\z,\y).%
\end{equation}
All aforementioned tie breakers can be shown to be coordinating. 
However, for example for binary strategies, the tie breaker of choosing the opposite strategy of one of your neighbors is not coordinating.  

There are several well-known update rules in network games for which various convergence and control results have been obtained \cite{optcontrol, ramazi2016networks}.
First, we present a brief introduction to some of the commonly appeared update rules in the literature and explore the conditions under which the resulting network games are coordinating. 
Next, we study networks with mixed dynamics where not all agents follow the same update rule.

\subsection{Myopic best-response} \label{sectionBestResponse}
Agents following the \textit{(myopic) best-response} update rule revise their strategies to the one that maximizes their total utility in  \eqref{eq:totalpayoff}.
Given the decision state $\bm{x}$, define $\x_{i = \k} \in \C^n$ to be the same as $\x$ but when agent $i$'s strategy is set to $\k$, i.e., $x_i = \k$.
Best response imposes the following feasible set for agent $i$:
\begin{equation}\label{eq:brdef}
    \mathcal{B}_i(\x) = 
    \left\lbrace \q \in \C \ |\ u_i(\x_{i = \q}) \geq u_i(\x_{i = \p})\ \forall \p \in \C \right\rbrace.
\end{equation}
So a \emph{best-responder} $i$ active at time $t$ updates to
\begin{equation*}
    x_i(t+1) \in \argmax_{\j \in \C}\ [\bm{\pi^i} \bm{y^i}(t)]_{\j},
\end{equation*}
where $[\bm{x}]_{\j}$ is the $\j^{\text{th}}$ entry of vector $\bm{x}$ and $\y^i\in \mathbb{Z}_{\geq0}^{|\C|}$ is the distribution of strategies among the neighbors of agent $i$ at state $\y$, i.e., $\bm{\y^i}_{\j}$ is the number of agent $i$'s neighbors who play strategy $\j$.

It was shown in \cite{control:bestresponse, riehl2018survey}, that under 2 available strategies, a best-responder is $\mathtt{A}$-coordinating, if her payoff matrix $\bm\pi$ satisfies 
\begin{equation*} 
    \pi_{1,1} - \pi_{2, 1} - \pi_{1, 2} + \pi_{2, 2}  > 0.
\end{equation*}
A special case of the above inequality is
\begin{equation} \label{BR_coordinating_binary}
    \pi_{\k,\k} > \pi_{\p,\k} \qquad \forall \k,\p\in\{\1,\2\}.
\end{equation}
Best-responders with payoff matrices satisfying \eqref{BR_coordinating_binary} are known as \emph{coordinating} or \emph{conforming best-responders} (or \emph{agents}), or simply \emph{coordinators} \cite{le2020heterogeneous} or \emph{conformists}.
Coordinating agents are shown to switch to a strategy if a sufficient number of their neighbors are playing that strategy -- a dynamic equivalent to the \emph{linear threshold model} \cite{ramazi2020convergence,granovetter1978threshold}.
Since the original definition of $\mathtt{A}$-coordinating populations does not take into account the role of tie breakers, we first provide the following result. 
\begin{lemma} \label{coordinatorsAreRestrictiveCoordinating}
    A coordinating best-responder with a coordinating tie breaker is restrictive coordinating for $|\C| = 2$. 
\end{lemma}
\begin{IEEEproof} 
    Consider coordinating agents $\N$ with the choice set $\C=\{\1,\2\}$.
    It suffices to prove that the population is $\1$-coordinating.
    Consider the decision states $\y,\z\in\C^n$ satisfying 
    \begin{equation}    \label{coordinatorsAreRestrictiveCoordinating-eq1}
        y_j = \1 \Rightarrow z_j = \1 \quad \forall j \in \N.
    \end{equation}
    If $\f_i(\y) = \tau_i(\mathcal{B}_i(\y),\y) = \1$, then either 
    $\mathcal{B}_i(\y) = \{\1\}$ or $\mathcal{B}_i(\y) = \{\1,\2\}$.
    In the first case, the following holds according to \eqref{eq:brdef}:
    \begin{equation}    \label{coordinatorsAreRestrictiveCoordinating-eq2}
        \pi^i_{\1,\1}y^i_\1 + \pi^i_{\1,\2}y^i_\2 
        > \pi^i_{\2,\1}y^i_\1 + \pi^i_{\2,\2}y^i_\2. 
    \end{equation}
    In view of \eqref{coordinatorsAreRestrictiveCoordinating-eq1}, 
    $
        z^i_\1 \geq y^i_\1 
    $,
    $
        z^i_\2 \leq y^i_\2
    $,
    and $z^i_\1 + z^i_\2$ = $y^i_\1 + y^i_\2$.
    Hence, \eqref{BR_coordinating_binary} implies
    \begin{equation*}
        \pi^i_{\1,\1}(z^i_\1-y^i_\1) > \pi^i_{\1,\2}(y^i_\2-z^i_\2)
        \Rightarrow
        \pi^i_{\1,\1}z^i_\1 + \pi^i_{\1,\2}z^i_\2 \geq
        \pi^i_{\1,\1}y^i_\1 + \pi^i_{\1,\2}y^i_\2 
    \end{equation*}
    and similarly,
    \begin{equation*}
        \pi^i_{\2,\1}y^i_\1 + \pi^i_{\2,\2}y^i_\2 \geq
         \pi^i_{\2,\1}z^i_\1 + \pi^i_{\2,\2}z^i_\2.
    \end{equation*}
    Therefore, in view of \eqref{coordinatorsAreRestrictiveCoordinating-eq2},
    \begin{equation*}    
        \pi^i_{\1,\1}z^i_\1 + \pi^i_{\1,\2}z^i_\2 
        > \pi^i_{\2,\1}z^i_\1 + \pi^i_{\2,\2}z^i_\2,
    \end{equation*}
    implying $\mathcal{B}_i(\z)=\{\1\}$, and hence, $\f_i(\z) = \tau_i(\mathcal{B}_i(\z),\z) = \1$.
    Now consider the case with $\mathcal{B}_i(\y) = \{\1,\2\}$.
    Similar to above, it can be shown that $\mathcal{B}_i(\z)$ is either $\{\1\}$ or $\{\1,\2\}$. 
    The first case readily results in $\f_i(\z) = \1$.
    So consider the case $\mathcal{B}_i(\z) = \{\1,\2\}$.
    Then since the tie breaker is coordinating, it holds that
    \begin{equation*}
        \tau_i(\mathcal{B}_i(\z),\z) \in \{\tau_i(\mathcal{B}_i(\z),\y)\} \cup \K(\z,\y).
    \end{equation*}
    On the other hand, 
    $\tau_i(\mathcal{B}_i(\z),\y) = \tau_i(\mathcal{B}_i(\y),\y) = \1$ and 
    $\K(\z,\y) = \{\1\}$.
    Thus, 
    \begin{equation*}
        \tau_i(\mathcal{B}_i(\z),\z) = \1,
    \end{equation*}
    resulting in again $\f_i(\z) = \1$.
    Hence, we showed that if $\f_i(\y) = \1$, then $\f_i(\z) = \1$, completing the proof.
\end{IEEEproof}

We can now extend the coordinating property to a population of coordinating agents. 
\begin{proposition} \label{coordinatorsAreRestrictiveCoordinating2}
    A network of coordinating best-responders with coordinating tie breakers is a restrictive coordinating population. 
\end{proposition}
\begin{IEEEproof}
    The proof follows Lemma \ref{coordinatorsAreRestrictiveCoordinating} and the definition of a restrictive coordinating population.
\end{IEEEproof}

The equilibration of the network game readily follows.
\begin{corollary}
    A network of coordinating best-responders with coordinating tie breakers almost surely equilibrates under any random activation sequence.
\end{corollary}
\begin{IEEEproof}
    The proof follows Theorem \ref{theorem_ACoordinatingEquilibrates} and Propositions \ref{restrictiveCoordinatingIsCoordinating} and \ref{coordinatorsAreRestrictiveCoordinating}.
\end{IEEEproof}

\begin{figure}
    \centering
    \begin{subfigure}[b]{0.22\textwidth}
        \includegraphics{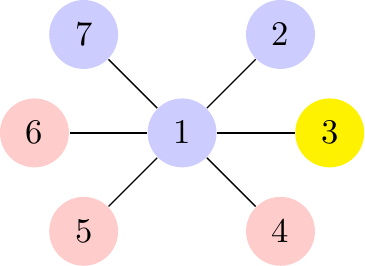}
        \caption{The population under state $\bm{y}$.}
    \end{subfigure}
    \vspace{5pt}
    \begin{subfigure}[b]{0.22\textwidth}
            \includegraphics{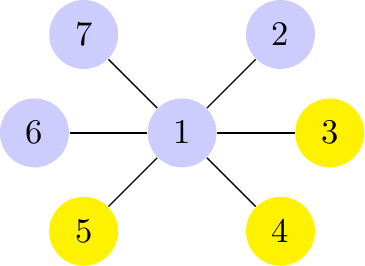}
        \caption{The population under state $\bm{z}$.}
    \end{subfigure}
    \caption{\textbf{The best-response network game in Example \ref{DiagonalIsNotRestrictiveCoordinating} under the two states $\y$ and  $\z$.} 
    The network game $((\N = \{1, 2, 3, 4, 5, 6, 7\}, \E), \bm{\pi}, \R$), is initially at state $\y = $ (\Bst, \Bst, \Yst, \Rst, \Rst, \Rst, \Bst) and the center agent tends to play \Bst. 
    But if the number of \Bst\ and \Yst\ agents both increase as in $\z$, agent 1 tends to play \Yst\, and the network cannot be \Bst\ or restrictive coordinating.}
    \label{fig:br_not_a_coord}
\end{figure}

\begin{figure}
    \centering
    \begin{subfigure}[b]{0.22\textwidth}
        \includegraphics{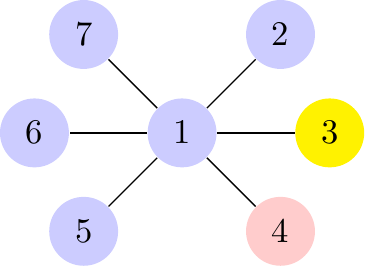}
        \caption{The population under state $\bm{y}$.}
    \end{subfigure}
    \vspace{5pt}
    \begin{subfigure}[b]{0.22\textwidth}
            \includegraphics{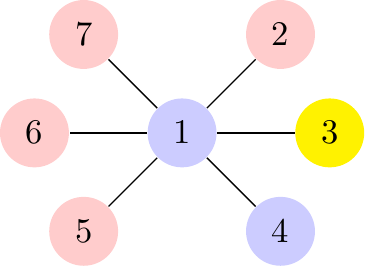}
        \caption{The population under state $\bm{z}$.}
    \end{subfigure}
    \caption{\textbf{The best-response network game in Example \ref{DiagonalIsNotCoordinating} under the two states $\y$ and  $\z$.} 
    The network game $((\N = \{1, 2, 3, 4, 5, 6, 7\}, \E), \bm{\pi}, \R$), is initially at state $\y = $ (\Bst, \Bst, \Yst, \Rst, \Bst, \Bst, \Bst, \Bst) and the center agent tends to play \Bst. 
    Even though $f_i(\y) =$ \Bst\  and $\K(\z, \y)=\{$\Bst, \Rst$\}$, agent 1 does not tend to play \Rst\  or \Bst\  in $\z$, and the population is not coordinating.}
    \label{fig:br_not_a_coord2}
\end{figure}

However, this does not readily extend to $|\C| \geq 3$. 
Even in the case where the payoff matrices are positive diagonal, which is perhaps the most straightforward extension to the above coordinating idea, the population is not necessarily restrictive coordinating or even coordinating. 
\begin{example} \label{DiagonalIsNotRestrictiveCoordinating}
    Consider the network game in Figure~\ref{fig:br_not_a_coord}, where 
    \[\bm{\pi}^1 = \begin{bmatrix}
    15 & 0 & 0 \\
    0 & 10 & 0 \\
    0 & 0 & 5
    \end{bmatrix}, \]
    where the first row corresponds to strategy \Yst, second to \Bst, and third to \Rst. 
    Agent $1$ tends to play \Bst\ at $\y$; however, she does not tend to do so at $\z$ despite the fact that $\y$ and $\z$ satisfy \eqref{ACoordinatingDefinitionEq1} for \Bst. 
    Hence, the network game is not \Bst-coordinating, nor is it restrictive coordinating.
    The intuitive reason is that although the number of \Bst-players has increased, so has the number of \Yst-players, which apparently had a higher effect on agent $1$'s tendency. 
    Note that there is not enough evidence to rule the population from being coordinating as $\K(\z, \y) = \{$\Bst, \Yst$\}$.
\end{example}
\begin{example} \label{DiagonalIsNotCoordinating}
    Consider the network game in Figure~\ref{fig:br_not_a_coord2}, where
    \[\bm{\pi}^1 = \begin{bmatrix}
    30 & 0 & 0 \\
    0 & 10 & 0 \\
    0 & 0 & 1
    \end{bmatrix}, \]
    where the first row corresponds to strategy \Yst, second to \Bst\ and third to \Rst. 
    Agent 4 who does not play \Bst\ at $\y$ plays \Bst\ at $\z$, resulting in \Bst\ $\in \K(\z,\y)$.
    However, agents 2, 5, 6 and 7 who play \Bst\ at $\y$ do not do so at $\z$.
    Hence, the total number of \Bst-playing neighbors of the central agent decreases, which changes her tendency to strategy \Yst\ that neither belongs to $\K(\z,\y) = \{$\Bst, \Rst$\}$ nor $\f_i(\y) =$ \Bst. 
    The population is, therefore, not coordinating. 
\end{example}



\subsection{Unconditional imitation}    \label{subsection_imitation}
The (unconditional) \textit{imitation} update rule dedicates the active agent to play the strategy of the highest-earning agent in its self-inclusive neighborhood:
\begin{equation}\label{eq:imdef}
    \mathcal{I}_{i}(\x) = \left\lbrace x_k \ |\ k \in \argmax_{j \in \bar{\N}_i} u_j(\x)\right\rbrace,
\end{equation}
where $\bar{\N}_i = \mathcal{N}_i \cup \{i\}$ is the self-inclusive neighborhood of agent $i$.
It is known that if the agents are \textit{opponent coordinating}, meaning that their payoff matrices satisfy the following condition:
\begin{equation}    \label{opponentCoordinating}
    \pi_{\k, \k} \geq \pi_{\k, \p} \quad \forall \k, \p \in \C,
\end{equation}
then the network equilibrates under any persistent, and even deterministic, activation sequence \cite{control:imitation, riehl2018survey}.
According to the condition, an opponent-coordinating imitator earns more payoff if her opponent plays the same strategy as she does.
It has been proven that binary network games consisting of opponent-coordinating imitators are $\mathtt{A}$-coordinating for any $\mathtt{A}\in\C$ \cite{control:imitation}.
Similar to Proposition \ref{coordinatorsAreRestrictiveCoordinating}, one can show the following result.
\begin{proposition} \label{prop:imcoo}
    A network of opponent-coordinating imitators with coordinating tie breakers is a restrictive coordinating population for $|\C|=2$. 
\end{proposition}
\begin{IEEEproof}
    The proof is similar to that of Proposition 3 and Proposition \ref{coordinatorsAreRestrictiveCoordinating} in \cite{riehl2017control}. 
\end{IEEEproof}

Interestingly, however, an opponent-coordinating agent is not necessarily coordinating (nor $\mathtt{A}$-coordinating).
The reason is that the feasible set of an imitator, i.e., \eqref{eq:imdef}, depends on other agents' utilities.
If the other agents are also opponent-coordinating, then the focal imitator will be coordinating, but the reverse is not always true. 
For example, suppose that the highest earning neighbor of an opponent-coordinating agent $i$ is playing $\A$ at state $\y$, and that by forcing some agents to switch to $\A$ to obtain the state $\z$, the highest earner's utility decreases and no longer remains highest. 
Instead, a new neighbor playing $\B$ earns most. 
Then $\K(\z,\y) = \A$ and $\f_i(\y)=\A$, but $\f_i(\z) = \mathtt{B}$, implying that agent $i$ is not coordinating. 

The following convergence result follows Theorem \ref{theorem_ACoordinatingEquilibrates}.
\begin{corollary}
    A network of opponent-coordinating imitators with coordinating tie breakers and $|\C|=2$ almost surely equilibrates under any random activation sequence.
\end{corollary}

\begin{figure}[ht]
\centering
    \begin{subfigure}[b]{0.37\textwidth}
            \includegraphics{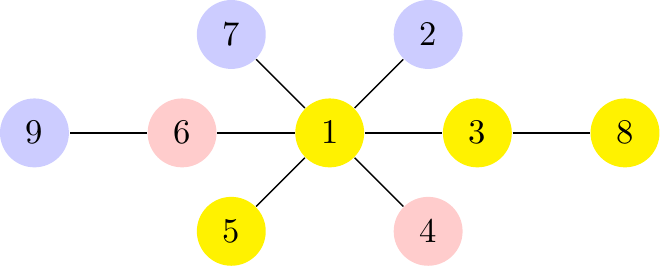}
    \caption{The population under state $\bm{y}$.}\label{fig:contim1}
    \end{subfigure}
    \\
    \vspace{15pt}
\begin{subfigure}[b]{0.37\textwidth}
            \includegraphics{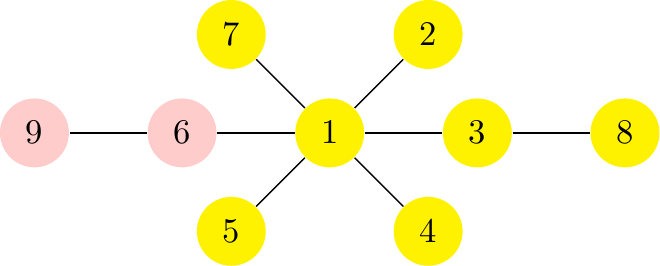}
    \caption{The population under state $\bm{z}$.}\label{fig:contim2}
    \end{subfigure}
    \caption{\textbf{An opponent-coordinating imitation network game shown under two states $\y$ and $\z$.} 
    The network game $(\mathbb{G} = (\N, \mathcal{E}), \bm{\pi}, \R)$ consists of $\N = \{1, 2, 3, 4, 5, 6, 7, 8, 9\}$, and three strategies, $\C = \{$\textcolor{red}{R}, \textcolor{orange}{Y}, \textcolor{blue}{B}$\}$.
    Even though that $\bm{y}, \bm{z}$ satisfy \eqref{ACoordinatingDefinitionEq1}, not all agents that want to play \textcolor{orange}{Y} under $\bm{y}$ still want to play \textcolor{orange}{Y} under $\bm{z}$, e.g., $\f_1(\bm{y}) =$ \textcolor{orange}{Y}, but $\f_1(\bm{z}) =$ \textcolor{red}{R}.}\label{fig:contim}
\end{figure}

For $|\C|\geq 3$, populations of opponent-coordinating agents may no longer be coordinating.
\begin{example}
    Consider the network game in \autoref{fig:contim}.
    It is clear that the two states represented by \autoref{fig:contim1} and \autoref{fig:contim2} satisfy \eqref{ACoordinatingDefinitionEq1} for the \textcolor{orange}{Y} strategy, i.e., all agents that play \textcolor{orange}{Y} under $\bm{y}$ also play \textcolor{orange}{Y} under $\bm{z}$. 
    Suppose that all agents have the same opponent-coordinating payoff matrix $\bm\pi$, and $\pi_{\text{\textcolor{red}{R}}, \text{\textcolor{red}{R}}} \gg \pi_{\text{\textcolor{orange}{Y}}, \text{\textcolor{orange}{Y}}}$.
    Then agent 1, that tends to play \textcolor{orange}{Y} under $\bm{y}$, tends to \textcolor{red}{R} under $\bm{z}$, because agent $6$ earns the highest payoff under $\bm{z}$.
    Therefore, this network is not restrictive coordinating. 
    On the other hand, $\K(\y,\z) = \{\text{\textcolor{blue}{B}}, \text{\textcolor{red}{R}}\}$. 
    Hence, $$\f_i(\y) = \text{\textcolor{orange}{Y}} \not\in \{\f_i(\z)\}\cup \K(\y,\z) = \{\text{\textcolor{blue}{B}}, \text{\textcolor{red}{R}}\}.$$
    This implies that the population is not coordinating either. 
\end{example}

\subsection{Rational imitation}\label{sec:ri}
The unconditional imitation update rule may lead agents to mimic a strategy even if it reduces their utilities. 
An intuitive alternative is to adopt the highest-earning strategy only if playing it does not decrease your utility, i.e., it is a better response to the neighbors.
This combination of imitation and best-response results in the so called \textit{rational imitation} \cite{govaert2021rationality} update rule, imposing the following feasible set for agent $i$:
\begin{equation}
    \mathcal{F}_i(\x) = 
    \begin{cases} 
    \mathcal{I}_i(\x) \cap \bar{\mathcal{B}}_i(\x) &\text{if  } \mathcal{I}_i(\x) \cap \bar{\mathcal{B}}_i(\x) \neq \emptyset\\
    \{x_i\} &\text{if  } \mathcal{I}_i(\x) \cap \bar{\mathcal{B}}_i(\x) = \emptyset
    \end{cases},
\end{equation}
where $\I_i$ is the imitation feasible set \eqref{eq:brdef}, and $\bar{\mathcal{B}}_i$ is the set of ``better replies'' consisting of all strategies that increase agent $i$'s utility:
\begin{equation*}
    \bar{\mathcal{B}}_i(\x) 
    = \left\lbrace \q \ |\ u_i(\x_{i = \q}) > u_i(\x)\right\rbrace.
\end{equation*}
It is not surprising for the coordinating condition in this case to be the intersection of the conditions for the best-responders and imitators.
A payoff matrix is that of a \emph{coordination game} or simply is a \emph{coordination payoff matrix} if it satisfies both both \eqref{BR_coordinating_binary} and \eqref{opponentCoordinating}, i.e., 
\begin{equation*}
    \pi_{\k,\k}\geq\pi_{\k,\q},\pi_{\q,\k}\qquad \forall \k,\q\in\C.
\end{equation*}
%
\begin{proposition} \label{prop:raimcoo}
    A network of rational imitators with coordinating tie breakers is a restrictive coordinating population for $|\C|=2$ if all agents have coordination payoff matrices. 
\end{proposition}
\begin{IEEEproof}
    Let $\C=\{\1,\2\}$.
    It suffices to prove that the population is $\1$-coordinating.
    Consider two states $\y,\z\in\C^n$ satisfying \eqref{ACoordinatingDefinitionEq1} for $\A = \1$, and let $i\in\N$.
    Either of the following cases is in force:
    
    \textit{Case 1:} $\mathcal{I}_i(\y) \cap \bar{\mathcal{B}}_i(\y) \neq \emptyset$. 
    Then $\bar{\mathcal{B}}_i(\y) \neq \emptyset$.
    Then similar to the proof of Lemma \ref{coordinatorsAreRestrictiveCoordinating} it can be shown that \eqref{ACoordinatingDefinitionEq2} holds for $\A = \1$.

    \textit{Case 2:} $\mathcal{I}_i(\y) \cap \bar{\mathcal{B}}_i(\y) = \emptyset$.
    Then by the definition of the ration imitation update rule, $\f_i(\y) = y_i$. 
    The equation $\mathcal{I}_i(\y) \cap \bar{\mathcal{B}}_i(\y) = \emptyset$ implies that either $\2\not\in\mathcal{I}_i(\y)$ or $\2\not\in\bar{\mathcal{B}}_i(\y)$.
    If $\2\not\in\mathcal{I}_i(\y)$, none of agent $i$'s highest-earning neighbors play $\2$ at $\y$. 
    Then the same holds at $\z$, because the payoff matrices satisfy \eqref{opponentCoordinating}, and hence, no $\2$-player's utility has increased at $\z$.
    Hence, $\2\not\in\mathcal{I}_i(\z)$.
    Now if $\2\not\in\bar{\mathcal{B}}_i(\y)$, then strategy $\1$ does not earn agent $i$ a higher utility than strategy $\2$ at $\y$. 
    Then the same holds at $\z$, because the payoff matrices satisfy \eqref{BR_coordinating_binary}, and hence, more neighbors playing $\1$ can only increase agent $i$'s utility. 
    Hence, $\2\not\in\bar{\mathcal{B}}_i(\z)$.
    Therefore, in both cases, $\2\not\in\mathcal{I}_i(\z) \cap \bar{\mathcal{B}}_i(\z)$.
    Hence, $\mathcal{I}_i(\z) \cap \bar{\mathcal{B}}_i(\z)$ either equals $\{\1\}$ or is empty.  
    The case with $\mathcal{I}_i(\z) \cap \bar{\mathcal{B}}_i(\z) = \{\1\}$ results in $\f_i(\z) = \1$, which in turn yields \eqref{ACoordinatingDefinitionEq2} for $\A=\1$.
    The case with $\mathcal{I}_i(\z) \cap \bar{\mathcal{B}}_i(\z) = \emptyset$ results in $\f_i(\z) = z_i$.
    On the other hand, in view of \eqref{ACoordinatingDefinitionEq1}, either $z_i = y_i$ or $z_i = \1$. 
    In either case, \eqref{ACoordinatingDefinitionEq2} is satisfied for $\A=\1$, completing the proof.
\end{IEEEproof}

\begin{corollary}
    A network of rational imitators with coordinating tie breakers and coordination payoff matrices almost surely equilibrates under any random activation sequence.
\end{corollary}
\begin{IEEEproof}
    The proof follows Theorem \ref{theorem_ACoordinatingEquilibrates} and Propositions \ref{restrictiveCoordinatingIsCoordinating} and \ref{prop:raimcoo}.
\end{IEEEproof}

\subsection{Network games with mixed update rules}
We now shift our focus to networks of agents following two or more update rules and obtain conditions for which they are coordinating. 
One natural extension to the convergence results on the different update rules in the previous subsections is to prove equilibration for a mixed population of coordinating best-responders, opponent-coordinating imitators, and coordinating rational imitators.
This, however, is not necessarily true. 
For example, as we discussed in Subsection \ref{subsection_imitation}, an opponent-coordinating imitator is not necessarily a coordinating agent -- it depends on the the other agents in the population. 
So we need to ensure that the mixture in question makes every type of agent coordinating.

\begin{proposition} \label{mixedPopulationIsRestrictiveCoordinating}
    A binary network game where each agent is associated with a coordination payoff matrix and follows one of the best-response, imitation, or rational imitation update rules, and uses a coordinating tie breaker is a restrictive coordinating population.
\end{proposition}
\begin{IEEEproof}
    The best-responders are restrictive coordinating in view of Lemma \ref{coordinatorsAreRestrictiveCoordinating}.
    Moreover, similar to the proof of Proposition \ref{prop:imcoo}, it can be shown that the imitators are restrictive coordinating, because the utility of the maximum earner in any agent's neighborhood increases as more agents play the same strategy as that of the maximum earner. 
    This is because a coordination payoff matrix satisfies \eqref{opponentCoordinating}.
    Finally, similar to the proof of Proposition \ref{prop:raimcoo}, it can be shown that the rational imitators are coordinating as well.
\end{IEEEproof}

Now without the use of any potential functions, we are able to provide the following convergence result on a wide class of general network games. 
\begin{theorem}[Equilibration of mixed network games]   \label{Th_mixedPopulationsEquilibrate}
    Consider a binary network game where each agent is associated with a coordination payoff matrix and follows one of the best-response, imitation, or rational imitation update rules, and uses a coordinating tie breaker. 
    The network game almost surely equilibrates under any random activation sequence.
\end{theorem}
\begin{IEEEproof}
    This result follows \autoref{theorem_ACoordinatingEquilibrates} and Proposition \ref{mixedPopulationIsRestrictiveCoordinating}.
\end{IEEEproof}

\section{Relation to supermodularity}   \label{sec_supermodularity}
Consider a game with players $\N$, ordered binary strategy set $\C = \{\mathtt{A},\mathtt{B}\}$ with the ordering $\mathtt{A}\succ\mathtt{B}$, and utility functions $u_i:\C^n\to\mathbb{R}, i\in\N$.
The utility $u_i$ is said to have \textit{increasing differences} or \textit{supermodularity} if for $\y$ and $\z$ satisfying Condition \eqref{ACoordinatingDefinitionEq1}, 
\begin{equation}    \label{supermodular}
    u_i(\z_{i=\A})-u_i(\z_{i=\mathtt{B}}) 
    \geq u_i(\y_{i=\A})-u_i(\y_{i=\mathtt{B}}),
\end{equation}
where $\x_{i=\X}$ is the same as $\x$ but where the $i^\text{th}$ component is set to $\X$ \cite{milgrom1990rationalizability, topkis1979equilibrium, durand2020optimal}.
For example, it is straightforward to show that a utility that is based on a coordination payoff matrix is supermodular. 

Under some mild conditions, a \textit{supermodular game} is defined based on the supermodular utilities and is proven to have certain properties, such as admitting a pure Nash equilibrium. 
The convergence results on supermodular games, however, are mainly limited to the best-response update rule \cite{chen2004does}.
A coordinating population is not necessarily a supermodular game, simply because a coordinating population may not even be a network game, i.e., the agents may not have utility functions. 
On the other hand, supermodular agents are not necessarily coordinating. 
The counter example is $u_i(\x) = 1- x_i$ with $\C=\{{1},{0}\}$. 
Because then \eqref{supermodular} is satisfied, but for $\y,\z\in\C^n$, where $z_i={1}$ and $y_i = {0}$, we have $f_i(\z)={0}$, $f_i(\y)={1}$, and $\Z(\z,\y) = \{{1}\}$, falsifying \eqref{coordinatingDefinitionEq-2}.

Nevertheless, the notion of supermodularity can be interpreted beyond utility functions. 
A reading that is in line with the so called \textit{strategic complementary} is that a player choosing ``a higher'' strategy, provides an incentive to all other players to also raise their choices, where ``highness'' is with respect to some partial ordering of the available choices \cite{bulow1985multimarket}. 
Let $s_i\in\{0,1\}$ denote agent $i$'s ``satisfaction,'' where $s_i = 0$ if and only if agent $i$ tends to switch. 
That is, 
\begin{equation*}
    s_i(\x) = 1 \iff x_i = f_i(\x).  
\end{equation*}
It is straightforward to show that \eqref{ACoordinatingDefinitionEq2} is equivalent to
\begin{equation*}    \label{supermodular2}
    s_i(\z_{i=\A})-s_i(\z_{i=\mathtt{B}}) 
    \geq s_i(\y_{i=\A})-s_i(\y_{i=\mathtt{B}}),
\end{equation*}
where $\y$ and $\z$ satisfy \eqref{ACoordinatingDefinitionEq1}.
Namely, restrictive coordinating agents' satisfaction is supermodular. 
Yet, as the utility is not necessarily supermodular, none of the established theories regarding supermodular games are applicable.

\section{A public goods game example}   \label{sec:publicGood}
In addition to pairwise interactions, the coordinating property is found in network games with group interactions.
In what follows, we define a \emph{spatial public goods game} \cite{govaert2021rationality}.
Consider the network $\mathbb{G} = (\N, \mathcal{E})$.
Each agent $i\in\N$ hosts a public goods game centered around herself and played with all of her neighbors.
So each agent participates in $|\bar{\N}_i|$ different games.
The agents choose from the strategy set $\C = \{$cooperate ($\mathtt{C}$), defect ($\mathtt{D}$)$\}$ and play the same strategy in all of the $|\bar{\N}_i|$ games.
In each game, a cooperator gives a fixed amount $c$ to the public pool and a defector gives nothing.
Afterward, The amount in the public pool gets multiplied by the \textit{public goods multiplier} $r > 0$, and then is distributed equally to all agents in the self-inclusive neighborhood.
Therefore, the payoff of agent $i$ from the public goods game centered around agent $j$ equals $\frac{rc(n^\Ce_j + x_j)}{|\bar{\N}_j|} - cx_i$, where $n^\Ce_j$ is the number of agent $j$'s cooperating neighbors, and
$x_j$ is the strategy of agent $j$, equal to 1 if she cooperates and 0 otherwise. 
Hence, agent $i$'s utility from all of the public goods game she participates in is
\begin{equation}
    u_i(\x) \triangleq \sum_{j \in \bar{\mathcal{N}}_i} \frac{rc(n^\Ce_j(\x) + x_j)}{|\bar{\mathcal{N}}_j|} - |\bar{\mathcal{N}}_i|cx_i.
\end{equation}
We investigate the behaviour of a spatial public goods game governed by the previous three update rules: the myopic best-response, unconditional imitation and rational imitation.

\subsection{Myopic Best-response}
Agents following the best-response update rule in public goods games revise their strategy based on \eqref{eq:brdef}.
That is, they maximize their utility against their neighbours' strategies. 

\begin{proposition}
    A spatial public goods game governed by the best-response update rule and coordinating tie-breakers is a restrictive coordinating population.
\end{proposition}
\begin{IEEEproof}
    Given a state $\x$, an agent $i\in\N$ tends to cooperate, i.e., $\f_i(\x) = \Ce$, only if
    \begin{align*}
         &\sum_{j \in \bar{\mathcal{N}}_i} \frac{rc(n^\Ce_j(\bm{x}) + x_j)}{|\bar{\mathcal{N}}_j|} - |\bar{\mathcal{N}}_i|c 
         \geq \sum_{j \in \bar{\mathcal{N}}_i} \frac{rc(n^\Ce_j(\bm{x}) + x_j)}{|\bar{\mathcal{N}}_j|}, \\
         \iff
         & \frac{rc(n^\Ce_i(\bm{x}) + 1)}{|\bar{\mathcal{N}}_i|} - |\bar{\mathcal{N}}_i|c \geq \frac{rcn^\Ce_i(\bm{x})}{|\bar{\mathcal{N}}_i|}\\
         \iff
         & r \geq |\bar{\N}_i|^2.
    \end{align*}
    Hence, her tendency towards cooperation is regardless of the decision state. 
    Therefore, if $\f_i(\y) = \Ce$, then because of having a coordinating tie-breaker, $\f_i(\z) = \Ce$ for any two states, including those satisfying \ref{ACoordinatingDefinitionEq1} for $\A = \Ce$. 
    So the agent is $\Ce$-coordinating. 
    Similarly, she is $\mathtt{D}$-coordinating. 
    Thus, the population is restrictive coordinating. 
\end{IEEEproof}

Consequently, we obtain the following convergence result which is in line with Corollary 1 in \cite{govaert2021rationality}.
\begin{corollary}
    A spatial public goods game governed by the best-response update rule and coordinating tie-breakers almost surely equilibrates under any random activation sequence. 
\end{corollary}
 
\subsection{Unconditional and rational imitation}
Active gents following the imitation update rule in public goods games choose their next strategy from \eqref{eq:imdef}.
Contrary to the similar case in the network games setting, imitators in public goods games are not necessarily \emph{$\mathtt{C}$-}coordinating, and do not reach an equilibrium.  
Examples of such networks are sketched in \cite{govaert2021rationality}.
 
\begin{figure}[ht]
    \centering
        \begin{subfigure}[b]{0.37\textwidth}
     \centering
    \includegraphics[width=0.8\linewidth]{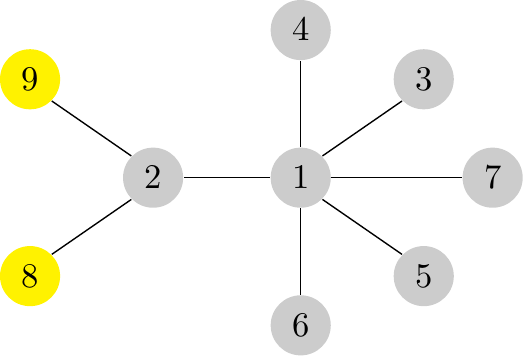}
        \caption{The population under state $\bm{y}$.}\label{fig:contim1pgg}
        \end{subfigure}
        \\
        \vspace{15pt}
    \begin{subfigure}[b]{0.37\textwidth}
     \centering
    \includegraphics[width=0.8\linewidth]{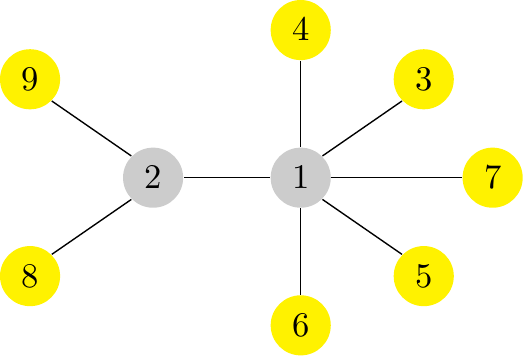}
        \caption{The population under state $\bm{z}$.}\label{fig:contim2pgg}
        \end{subfigure}
    \caption{\textbf{A spatial public goods game governed by the imitation update rule, under two states $\y$ and $\z$.}
    The colors yellow and gray represent cooperation ($\Ce$) and defection ($\mathtt{D}$).
    Agent 2 tends to cooperate under $\y$ as her highest-earning neighbours are 8 and 9, but tends to defect under $\z$ as agent 1 is the highest-earner in her neighborhood.
    Given that $\K(\z,\y) = \{\Ce\}$, $f_2(\y) = \Ce$, but $f_2(\z) = \mathtt{D}$, the population is not coordinating.}\label{fig:contimpgg}
\end{figure}

Since rational imitation does not blindly replicate the strategy of the highest-earning agent, one may hypothesize that it leads the game to equilibrium.
It has been proven \cite{govaert2021rationality} that, indeed, this is the case, i.e., a spatial public goods game equilibrates under the imitation update rule.
Nevertheless, the population may not be coordinating. 
This is a result of some of agent $i$'s neighbors having a larger set of coordinating neighbors and as a result, accumulating more payoff (Figure~\ref{fig:contimpgg}).

\section{Control}   \label{sec_control}
It is often desired to drive the decision-making dynamics from one equilibrium to another where a particular choice, say $\mathtt{X}\in\C$, is more frequent. 
This is typically done by assuming that a central agency has the power to modify the revision protocol of a number of agents by means of providing incentives, so that they tend to choose $\mathtt{X}$ instead of their current choice.
As these agents update their choices to $\mathtt{X}$, other agents may also switch to $\mathtt{X}$.
How many agents must be incentified to ensure a certain number of $\mathtt{X}$-selectors is reached in the population?
In general, the answer depends on the activation sequence decision-making dynamics, because starting from the same initial condition, the dynamics may reach different equilibria under different activation sequences. 
This is, however, not the case for coordinating populations, as explained in the following. 

\subsection{Convergence under incentive}\label{sec:cui}
We say that agent $i$ is \textit{provided incentive (reward) $r$ to choose $\mathtt{X}\in\C$ at state $\x\in\C^n$} if her revision protocol is changed to $f^r_i$, where $f^r_i(\x) = \mathtt{X}$ if the incentive is sufficient to change her tendency towards $\mathtt{X}$, and $f^r_i(\x) = f_i(\x)$ if it is insufficient.
In case an agent is not provided incentive ($r=0$), we define $f^r_i(\x) = f_i(\x)$.
First, we show that for every choice $\mathtt{X}\in\C$, coordinating populations are \textit{monotone in $\mathtt{X}$}; that is, after offering incentives to agents for choosing $\mathtt{X}$ at equilibrium, no agent will switch away from $\mathtt{X}$ \cite{riehl2018survey}.
\begin{lemma}[Monotone in $\mathtt{X}$]   \label{lem_monotoneInX}
    Consider a coordinating population defined by $(\N,\C,\F)$ that is at equilibrium $\x^*$. 
    If some of the agents are provided incentive to choose an arbitrary choice $\mathtt{X}\in\C$ at $\x^*$, then the agents may switch to only $\mathtt{X}$ under any activation sequence. 
\end{lemma}
\begin{IEEEproof}
    We prove by induction on time: At every time $t\in\mathbb{Z}_{\geq0}$, the active agent may switch to only $\mathtt{X}$.
    The result is trivial for $t=0$. 
    Assume that the inductive statement holds for $t=0,1,\ldots, T$. 
    Consider the states $\y = \x^*$ and $\z = \x(T)$. 
    It holds that $\K(\z,\y) = \{\mathtt{X}\}$. 
    Hence, as the population is coordinating, \eqref{coordinatingDefinitionEq-2} implies that for every agent, including the one active at time $T$, denoted by $i$, it holds that $\f^r_i(\z)$ equals $\mathtt{X}$ or $f^r_i(\y)$, where $r$ is the reward that agent $i$ received.
    The first case implies that agent $i$ switches to $\mathtt{X}$ at time $T+1$. 
    For the second case, if agent $i$ was one of the incentified agents at $\y = \x^*$, then $f^r_i(\y) = \mathtt{X}$ by definition; otherwise, she was satisfied with her choice as $\x^*$ is an equilibrium, yielding $f^r_i(\y) = y_i$.
    So agent $i$ will either switch to $\mathtt{X}$ or stick to her previous choice at time $T+1$.
    This completes the inductive step and the proof. 
\end{IEEEproof}

Next, we discuss the so called \textit{uniquely convergent in $\mathtt{X}$} property \cite{riehl2018survey, riehl2017control}; that is, after offering incentives to agents for choosing $\mathtt{X}$ at equilibrium, the population will reach a unique equilibrium under any \textit{persistent} activation sequence, where every agent becomes active infinitely many times \cite{ramazi2016networks}.
The persistent activation assumption is weaker than the random activation assumption as every random sequence is also persistent but not vice versa. 
If the population is uniquely convergent in $\mathtt{X}$ for every choice $\mathtt{X}\in\C$, we say it is \textit{uniquely convergent}. 

\begin{proposition}[Uniquely convergent] \label{prop_uniqueConvergence}
    Consider a coordinating population defined by $(\N,\C,\F)$ that is at equilibrium $\x^*$. 
    If some of the agents are incentified to choose $\mathtt{X}\in\C$ at $\x^*$, then the decision-making dynamics will reach a unique equilibrium under a persistent activation sequence. 
\end{proposition}
\begin{IEEEproof}
    In view of Lemma \ref{lem_monotoneInX}, the agents can only switch to $\X$. 
    However, they can only make finite switches, so they will stop switching at some state, which has to be an equilibrium due to the persistent activation assumption. 
    We prove by contradiction that the reached equilibrium is unique. 
    Assume on the contrary that two different equilibria $\bm a^*$ and $\bm b^*$ are reached under activation sequences $\mathcal{A}$ and $\mathcal{B}$.  
    Let $i\in\N$ be an agent who has different choices at $\bm a^*$ and $\bm b^*$.
    According to Lemma \ref{lem_monotoneInX}, her choice has to be $\mathtt{X}$ at one of them, say $\bm a^*$, and $\x^*_i\neq\X$ at the other, i.e., $\bm{a}^*_i = \mathtt{X}$ and $\bm{b}^*_i = \x^*_i$.
    In view of Lemma \ref{lem_monotoneInX}, the network is monotone in $\X$, implying that no agent will switch away from $\X$ under either sequence. 
    So agent $i$ never switches to $\X$ under $\B$.
    Now let $j\in\N$ denote the first agent that switches to $\X$ under $\mathcal{A}$ but never switches to $\X$ under $\B$.
    Such an agent does not exist, because upon activation under $\B$, the state is the same as upon activation under $\mathcal{A}$ with the difference that some agents may have switched to $\X$ under $\B$, which forces the agent to choose $\X$ as the population is coordinating.
    Therefore, every agent, including agent $i$, who switches to $\X$ under $\mathcal{A}$ will also switch to $\X$ under $\B$. 
    This is a contradiction, leading to the proof.
\end{IEEEproof}

In what follows, we often focus on the special case presented in Section \ref{sec:netgames}, binary network games, particularly under the best-response and imitation update rules. 
Without loss of generality, we consider strategy $\1$ as the desired strategy.
We assume a central agency has the resources to reward any agent $i$ to play strategy $\1$.
That is,
\begin{equation}\label{eq:incentive}
    \bm{\hat{\pi}^i} \triangleq \begin{pmatrix}
    \pi_{\1,\1} + r_i & \pi_{\1,\2}+r_i\\
    \pi_{\2,\1} & \pi_{\2,\2}\\
    \end{pmatrix}, \quad r_i\geq0,
\end{equation}
where $r_i$ is the \textit{payoff incentive} to agent $i$ for playing $\1$.
We proceed to the three control problems introduced in \cite{riehl2018incentive, riehl2017control}.

\subsection{Uniform Reward Control}\label{sec:urc}
Suppose that at some equilibrium state $\x^*$, the central agency offers the same reward $r$ to all agents for playing strategy $\X\in\C$. 
\begin{problem}[Uniform reward control] \label{prob:uniform}
    Given the decision-making population $(\N,\C,\F)$ and initial equilibrium state $\x(0)=\x^*$, find the minimum reward $r^*$, where for all $r > r^*$, every agent will eventually fix her strategy to $\X$.
\end{problem}
The existence of $r^*$ depends on the update rule and initial state.
For example, in a network of imitators that all play strategy $\2$, none of the agents can be forced to play strategy $\1$, regardless of the amount of incentive. 
In \cite{riehl2018survey}, finite sets of candidate rewards $\mathcal{R}_I$ and $\mathcal{R}_B$ are defined for networks of coordinating best-responders and networks of opponent-coordinating imitators, respectively.
These sets include the optimal reward $r^*$, which is found by performing a binary search over the sets \cite{control:bestresponse,control:imitation}.
Motivated by the idea, we define the \emph{candidate reward set} for an agent $i$ following revision protocol $\f_i$ and at equilibrium state $\x^*$ as
\begin{align*}
    &\mathcal{R}_i(\x^*) = \\
    &\left\{\inf\{r \,|\, \f^r_i(\x) = \X\}\,|\, \x\in\C^n, x_i = \x^*_i, x_j \in\{\x^*_j,\X\}\forall j\neq i \right\}.
\end{align*}
Each reward in $\mathcal{R}_i$ will make agent $i$ tend to choose $\X$ at a particular state $\x$ where she has chosen her strategy at $\x^*$.
\begin{proposition} \label{optimalUniformRewardIsInR}
    If $r^*$ exists for a coordinating population, then $r^*\in\cup_{i\in\N}\mathcal{R}_i(\x^*)$.
\end{proposition}
\begin{IEEEproof}
    The existence of $r^*$ implies that for every $r>r^*$, all agents will fix their strategies to $\X$ under any activation sequence, and for every $r<r^*$, at least one agent will not fix her strategy to $\X$ under any activation sequence. 
    Hence, according to Proposition \ref{prop_uniqueConvergence}, there exists an agent $i\in\N$ who does not switch to $\X$ for any $r<r^*$, but does so for any $r>r^*$, under every activation sequence.
    Hence, $r^*$ is the infimum reward that ensures for all $r>r^*$, agent $i$ will switch to $\X$ at some (not necessarily unique) state $\x$ under any activation sequence.
    That is, $r^* = \inf\{r \,|\, \f^r_i(\x) = \X\}$, for all states $\x$ that may appear under different activation sequences and for all agents $i$ who do not switch to $\X$ for $r<r^*$ but do so otherwise.
    In view of Lemma \ref{lem_monotoneInX}, for every agent $j\in\N$, $x_j$ will be either $\x^*_j$ or $\X$ at any such state $\x$. 
    The proof is complete since all agents $i$ and such states $\x$ are covered in the definition of $\mathcal{R}_i$. 
\end{IEEEproof}

For binary decisions, i.e., $|\C|=2$, 
Proposition \ref{optimalUniformRewardIsInR} allows us to find the optimal reward by performing a binary search over $\cup_{i\in\N}\mathcal{R}_i$ and simulating the population using an arbitrary persistent activation sequence until it reaches an equilibrium. 
The full algorithm is presented in \cite{control:bestresponse}.

Now we generalize the results in \cite{control:bestresponse} and \cite{control:imitation} to a mixed network of best-responders and imitators. 
\begin{corollary}
    Consider the binary network game $(\mathbb{G}, \bm{\pi}, \F)$, where a non-empty subset of the agents $\B \subseteq \N$ are best-responders and agents in $\mathcal{I}=\N\backslash \B$ are imitators.
    Moreover, all agents are associated with a coordination payoff matrix and use a coordinating tie breaker. 
    Then $r^*\in \mathcal{R}_{B}\cup\mathcal{R}_{I}$.
\end{corollary}
\begin{IEEEproof}
    The existence of $r^*$ for any network of best-responders was proven in \cite{control:bestresponse}.
    Since the network has at least one best-responder, there exists some $r^*$ that forces her to play $\1$. 
    On the other hand, the existence of $r^*$ for any network of opponent-coordinating imitators was proven in \cite{control:imitation}, on the condition that the initial state is not $(\2)_{i = 1}^n$.
    Hence, $r^*$ exists. 
    It can be shown that $\mathcal{R}_{B} \subseteq \cup_{i\in\B} \mathcal{R}_i$ and $\mathcal{R}_{I} \subseteq \cup_{i\in\I} \mathcal{R}_i$.
    The proof then follows Proposition \ref{optimalUniformRewardIsInR}.
\end{IEEEproof}


\subsection{Targeted Reward Control}
Intuitively, if the incentives are tuned to each agent separately, a lower overall incentive is required to drive the population to the desired state.
This motivates studying the following problem.
\begin{problem}[Targeted reward control] \label{prob:targeted}
    Given the decision-making population $(\N,\C,\F)$ and initial equilibrium state $\x(0)=\x^*$, find the targeted reward vector $\bm{r}^* = (r^*_1, \dots, r^*_n)$ that minimizes $\sum_{i \in \N} r^*_i$ such that if $r_i > r^*_i$ for all $i \in\N$, every agent will eventually fix her strategy to $\X$.
\end{problem}
In targeted rewarding, the problem is more about which agents to be incentified and in what order, rather than how much the incentives should be. 
That is, if an agent $i$ is chosen to be rewarded, her infimum reward $r^*_i$ can be calculated based on the decision state.
For example, as discussed in Section \ref{sectionBestResponse}, a coordinating best-response agent $i$ plays $\1$ if the number of her $\1$-playing neighbors is more than some threshold, i.e., $(\gamma_i/\delta_i)|\N_i|$, where $\gamma_i = \pi^i_{\2,\2} - \pi^i_{\1,\2}$ and $\delta_i = \pi_{1,1} - \pi_{2, 1} - \pi_{1, 2} + \pi_{2, 2}$.
To make agent $i$ switch to strategy $\1$ at state $\x$, her threshold must fall short of the number of her $\1$-playing neighbors.
This results in
\begin{equation}\label{eq:brri}
    r^*_i = \gamma_i - \frac{\delta_i n_i^\1(\x)}{|\N_i|},
\end{equation}
where $n_i^{\1}(\x)$ denotes the number of $\1$-playing neighbors of agent $i$ \cite{riehl2018survey}.
The same calculation is more complicated for imitators.
In order to make agent $j$ switch to strategy $\1$, there should be a $\1$-playing agent $i$ among her neighbors, so that by providing incentive to agent $i$ and increasing her payoff, agent $j$ will tend to imitate agent $i$.
Thus, the payoff of agent $i$ has to exceed that of the highest-earning $\2$-player in agent $j$'s neighborhood:
\begin{equation}\label{eq:imri}
    r^*_i 
    = \max_{k \in \mathcal{N}_j^{\mathtt{2}}} u_k(\x) - u_i(\x), 
\end{equation}
where $\bar{\N}_j^{\2} = \{k \in \bar{\N}_j \ |\ x_k = \2\}$ is the set of $\2$-playing agents in agent $i$'s self-inclusive neighborhood.
\begin{algorithm}[!hb]
\caption{Generic iterative algorithm that solves Problem \ref{prob:targeted} for mixed binary networks of best-responders and imitators associated with coordination payoff matrices and coordinating tie breakers.
    The variable $\epsilon$ is an arbitrary small positive constant to ensure that the reward exceeds its infimum.}
    \label{alg:targeted}
    \begin{algorithmic}[1]
        \State Initialize $\x = \x(0)$ and $r_i = 0$ for every agent $i\in\N$ 
        \While{$\exists i \in \N: x_i = \2$}
            \State $\B := \{i \in \N \,|\, \Fe_i = \B_i \wedge x_i = \2\}$
            \State $\I := \{i \in \N\,|\, \Fe_i = \I_i \wedge x_i = \1 \wedge (\exists j \in \mathcal{N}_i : x_j = \2)\}$
            \State Choose agent $i$ from $\B \cup \I$ 
            \If{$\Fe_i=\B_i$}
                \State $r^*_i := \gamma_i - \frac{\delta_i n_i^\1(\x)}{|\N_i|}$ 
            \Else
                \State $r^*_i := \min_{j \in \N_i^{\2}} \max_{k \in \bar{\N}_j^{\2}} u_k(\x) - u_i(\x)$
            \EndIf
            \State $r_i := r_i + r^*_i + \epsilon$ 
            \State ${\Gamma} := (\mathbb{G}, \bm{\hat{\pi}}, \F)$
            \State Simulate ${\Gamma}$ until equilibrium $\bm{\hat{x}}$ is reached
            \State $\x := \bm{\hat{x}}$
        \EndWhile
        \State \textbf{return} $\bm{r}$
    \end{algorithmic}
\end{algorithm}

Now consider a mixed network of best-responders and imitators associated with coordination payoff matrices and using coordinating tie breakers. 
This algorithm provides a general framework for testing different methods of finding (approximations of) $r^*$.
In particular, each method of finding $r^*$ specifies how the agent to be incentified is chosen; that is, which agent $i$ will be chosen from $\mathcal{B}\cup\mathcal{I}$ in Line 5. 
As the network game is uniquely convergent in view of Theorem \ref{Th_mixedPopulationsEquilibrate}, giving an incentive to one agent will lead to a unique equilibrium independent of the activation sequence chosen in the simulation.
Therefore, the optimal amount of incentives is obtainable by searching exhaustively through all permutations of agents chosen in line 5 of the algorithm.
As this approach is not practical on large networks, there are several heuristics designed to tackle the complexity of the algorithm.

Proposed in \cite{control:bestresponse, control:imitation}, the \textit{iterative potential to reward optimization (IPRO)} heuristic uses a \textit{potential function} $\Phi$ that increases when an agent switches to the desired strategy, reaching its unique maximum at the state where all agents play the desired strategy.
Given the state $\x$, let $\bm{\hat{x}}$ be the equilibrium resulting from offering incentive to agent $j$. 
IPRO is defined as Algorithm \ref{alg:targeted} where in step 5, the agent is chosen by
\begin{equation}
    \argmax_{j\in \N}\frac{\Phi(\bm{\hat{x}}) - \Phi(\x)}{r^\beta_j},
\end{equation}
where $\gamma \geq 0$ are hyper-parameters\footnote{The original objective function in \cite{control:bestresponse} is  $\frac{(\Phi(\bm{\hat{x}}) - \Phi(\x))^\alpha}{r^\gamma_j}$ for $\alpha,\gamma\geq0$, which can be simplified as above under the $\argmax$ operator.}. 

IPRO requires the potential function $\Phi$ to be known.
If such a potential function is not found for the specific update rule in question, there will be no IPRO algorithm. 
Indeed, IPRO has only been defined for networks of coordinating best-responders \cite{riehl2018incentive} and networks of opponent-coordinating imitators \cite{riehl2017control,control:imitation}, because only for these two networks, a potential function has been found. 
The potential function for coordinating best-responders equals $\sum_{i\in\N} \Phi_i(\x)$, in which
 \begin{equation*}
    \Phi_i({\x}) = \begin{cases} n_i^\1(\x) - {n}^\tau_i & x_i = \1 \\
    n_i^\1(\x) - {n}^\tau_i + 1 &x_i = \2
    \end{cases},
\end{equation*}
with ${n}^\tau_i = \ceil{(\gamma_i/\delta_i)|\N_i|}$ being the minimum number of agent $i$'s $\1$-playing neighbors required for her to switch or continue playing $\1$.
The analogous function in an opponent-coordinating imitation network game is defined by
\begin{equation*}
    \Phi({\x}) = \sum_{i=1}^n n_i^\1({\x}),
\end{equation*}
which increases as the number of $\1$-players increases and is maximized when all agents play $\1$.

As there has not been any potential function introduced for mixed networks of best-responders and imitators, IPRO may not be used in this case.
By simply tracking the change in $n^\X$, the number of agents choosing the desired strategy $\X$, we propose a new algorithm, the \emph{iterative number to reward optimization (INRO) algorithm} that can be applied to any decision-making population, without the need of a potential function.
INRO is defined as Algorithm \ref{alg:targeted} where agent $i$ at line 5 is chosen from 
\begin{equation*}
    \argmax_{j\in \N}\frac{
        n^{\X}(\bm{\hat{x}})-n^{\X}(\bm{x})
    }{r^\beta_j}.
\end{equation*}

\begin{remark}
    The computational complexity of Algorithm \ref{alg:targeted} using either IPRO or INRO is at most $O(n^3)$. 
    The number of iterations for the main loop will be at most $n$, in the worst case offering incentives to all agents. 
    On the other hand, choosing an agent in each iteration takes at most $O(n^2)$ computational steps by activating all agents $n$ times, which guarantees equilibration according to Lemma \ref{lem_monotoneInX}. 
\end{remark}

\subsection{Budgeted Targeted Reward Control}
In many real-life situations, the central agency has a limited incentive budget, leading to the following problem.
\begin{problem}\label{prob:budgeted}
     Given the decision-making population $(\N,\C,\F)$, initial equilibrium state $\x(0)=\x^*$, and budget constraint $\sum_{i \in \N} r_i \leq \rho$, find the targeted reward vector $\bm{r^*}$ that maximizes the number of agents who fix their choices to $\X$.
\end{problem}
One may modify Algorithm \ref{alg:targeted} to account for the budget constraint, by limiting agent $i$ at line 5 to those agents whose required incentive $r^*_i$ does not violate the budget limit, and terminating the algorithm if there is no other agent that can be switched without exceeding the limit.
Both IPRO and INRO trivially extend to Problem \ref{prob:budgeted}. 

\subsection{Simulations} \label{secsims}
We test and compare the performance of INRO against the following previously proposed control algorithms \cite{control:bestresponse, control:imitation} for networks of best-responders, networks of imitators, and mixed networks of best-responders and imitators:
\begin{itemize}
    \item \emph{iterative degree:} targets the agent with maximum degree.
    \item \emph{iterative random:} targets a random agent.
    \item \emph{IPRO:} targets agents maximizing the potential-change-to-reward ratio, with $\beta = 4$.
    \item \emph{Optimal:} Preforms exhaustive search to find the optimal solution.
\end{itemize}

For all of the simulations, networks where randomly generated with connection radius $\sqrt{(1 + \mathtt{E[deg]})/n}$ where $\mathtt{E[deg]}$ is the expected degree of the nodes. 
The payoffs for the best-response agents were generated so that the thresholds are uniformly distributed on the interval $[0, \frac{2}{3}]$.
Similarly, for the imitators, payoffs were randomly generated by
$\bm{\pi}^i = p_iI + v_iW_i$, where $p_i \geq 1$ is the coordination level, forcing the payoff matrix to be that of a coordination game, $I$ is the $2\times2$ identity matrix, $v_i \in [0, 1]$ is the payoff variance, and $W_i$ is a $2 \times 2$ random matrix whose elements are generated independently at random from a uniform distribution on $[0, 1]$.

The initial state of the generated best-response and mixed networks is one where every agent is playing the undesirable strategy. 
The initial state for the only-imitator networks is chosen at random, with the probability of each agent playing the desirable strategy is $\frac{1}{3}$, as an imitator-only network that starts in a state where everyone is playing the undesirable strategy cannot be controlled using incentives.

\begin{figure}[ht]
\centering
    \begin{subfigure}[b]{0.45\textwidth}
            \includegraphics[width=\linewidth]{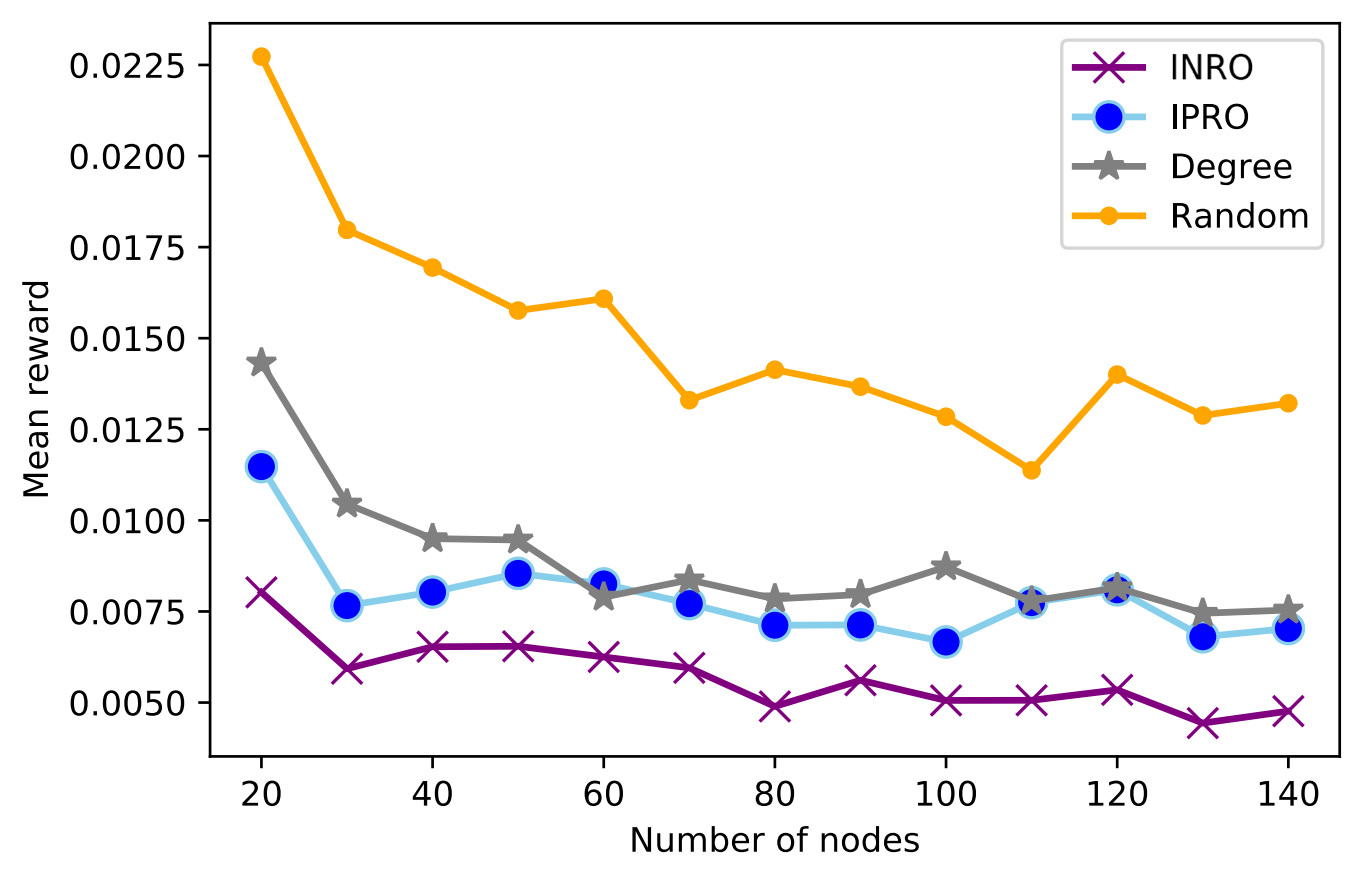}
    \caption{Performance comparison between four methods for Algorithm \ref{alg:targeted} for populations of different sizes on a network of coordinating best-responders.}\label{fig:sim_br1}
    \end{subfigure}
    \\
    \vspace{15pt}
\begin{subfigure}[b]{0.45\textwidth}
            \includegraphics[width=\linewidth]{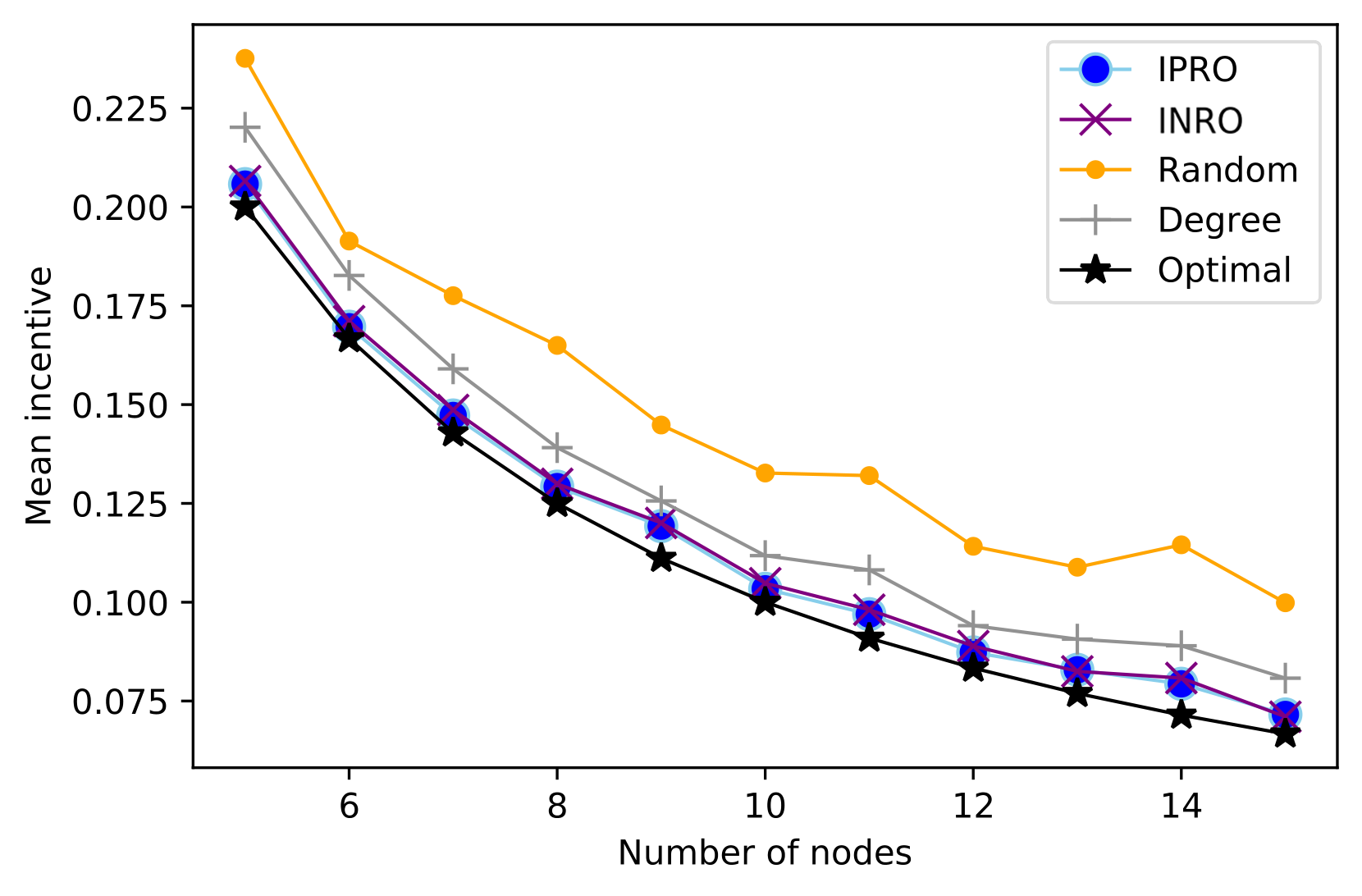}
    \caption{Performance comparison between five methods for Algorithm \ref{alg:targeted} for populations of different sizes on a network of coordinating best-responders.}\label{fig:sim_br2}
    \end{subfigure}
    \\
    \begin{subfigure}[b]{0.45\textwidth}
            \includegraphics[width=\linewidth]{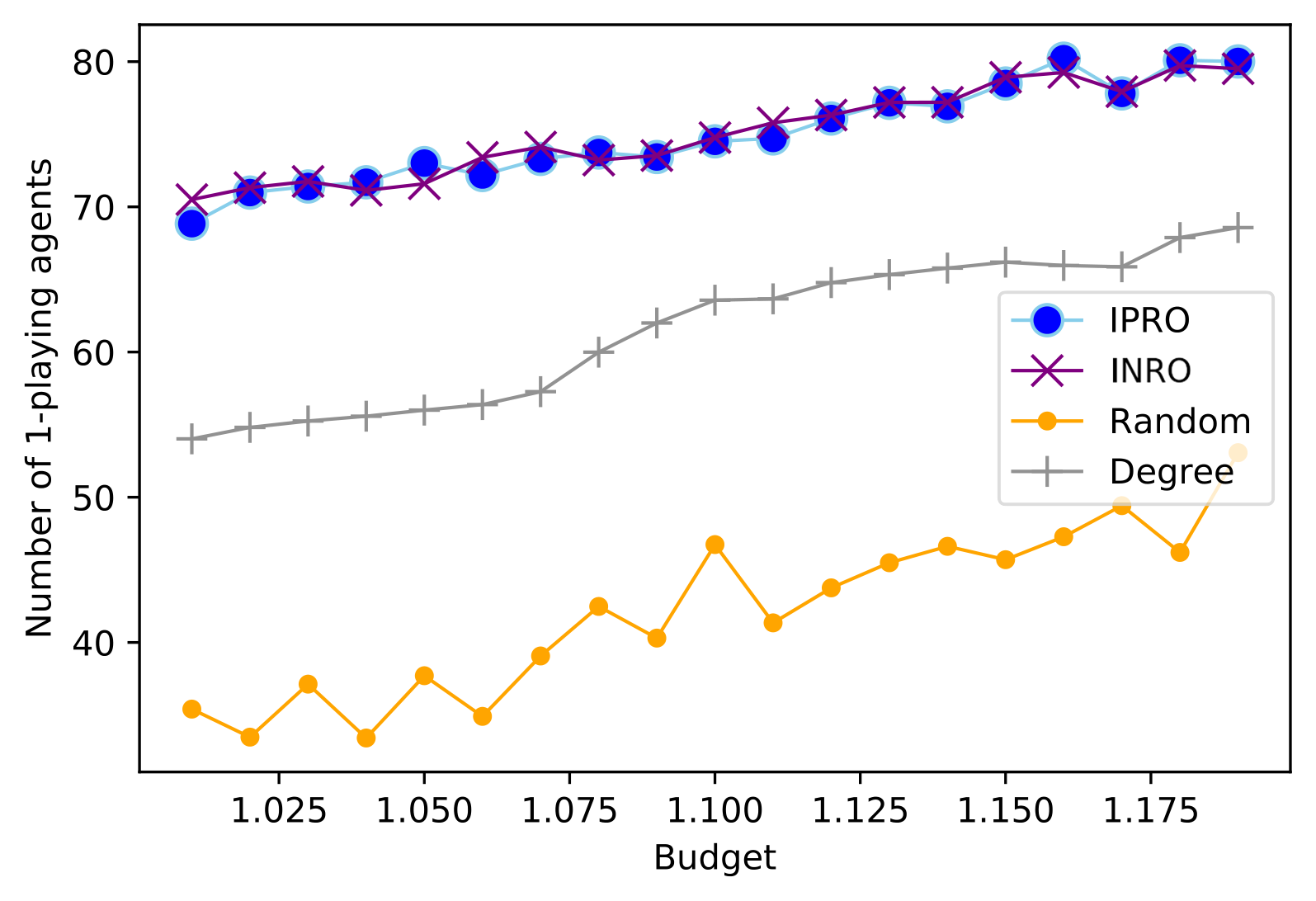}
    \caption{Performance comparison between four methods in solving Problem~\ref{prob:budgeted} for populations of size 100 on a network of coordinating best response agents.}\label{fig:sim_br3}
    \end{subfigure}
    \caption{\textbf{Targeted reward control of networks of coordinating best-responders.}}\label{fig:sim_br}
\end{figure}

\begin{figure}[ht]
\centering
    \begin{subfigure}[b]{0.45\textwidth}
            \includegraphics[width=\linewidth]{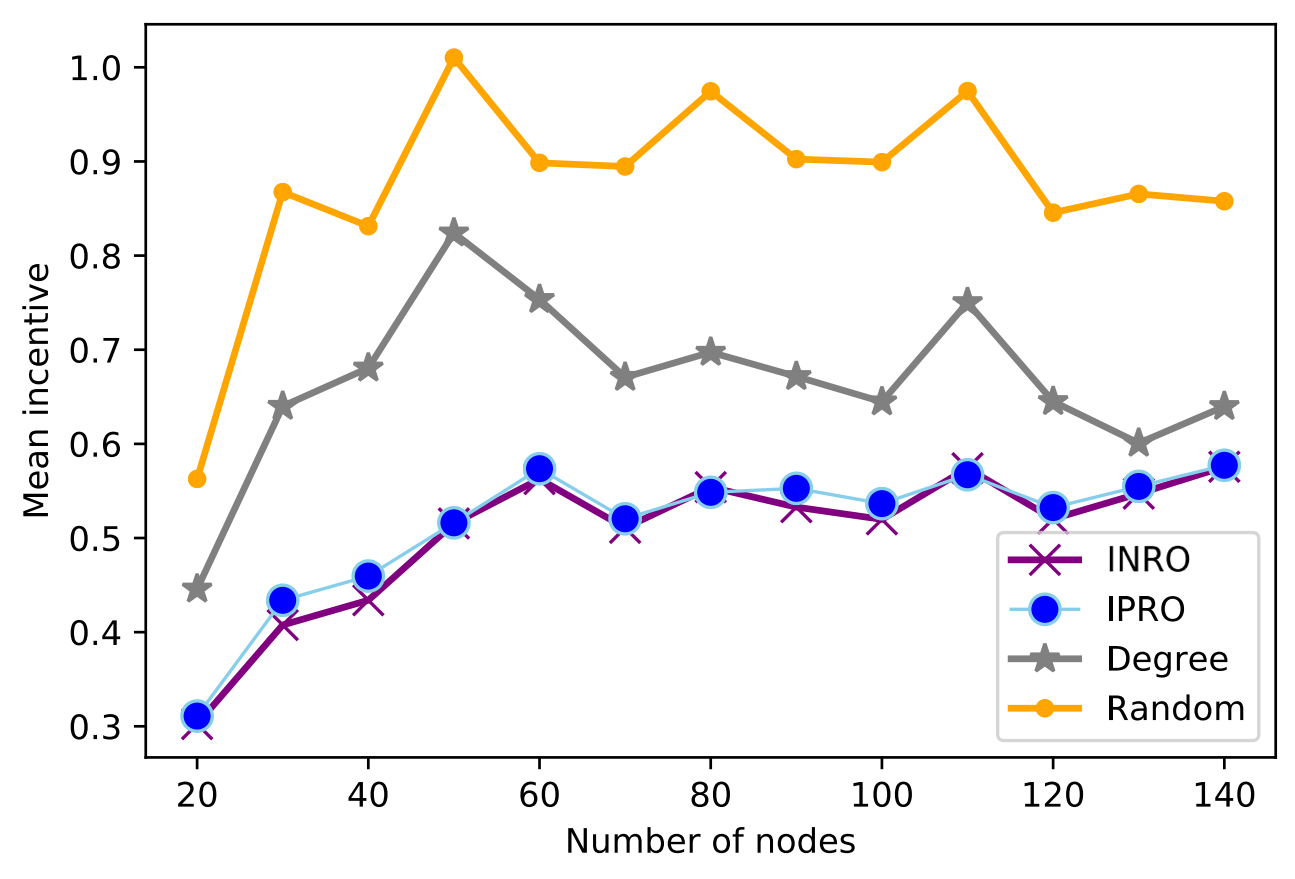}
    \caption{Performance comparison between four methods for Algorithm \ref{alg:targeted} for populations of different sizes on a network of opponent-coordinating imitators.}\label{fig:sim_im1}
    \end{subfigure}
    \\
    \vspace{15pt}
\begin{subfigure}[b]{0.45\textwidth}
            \includegraphics[width=\linewidth]{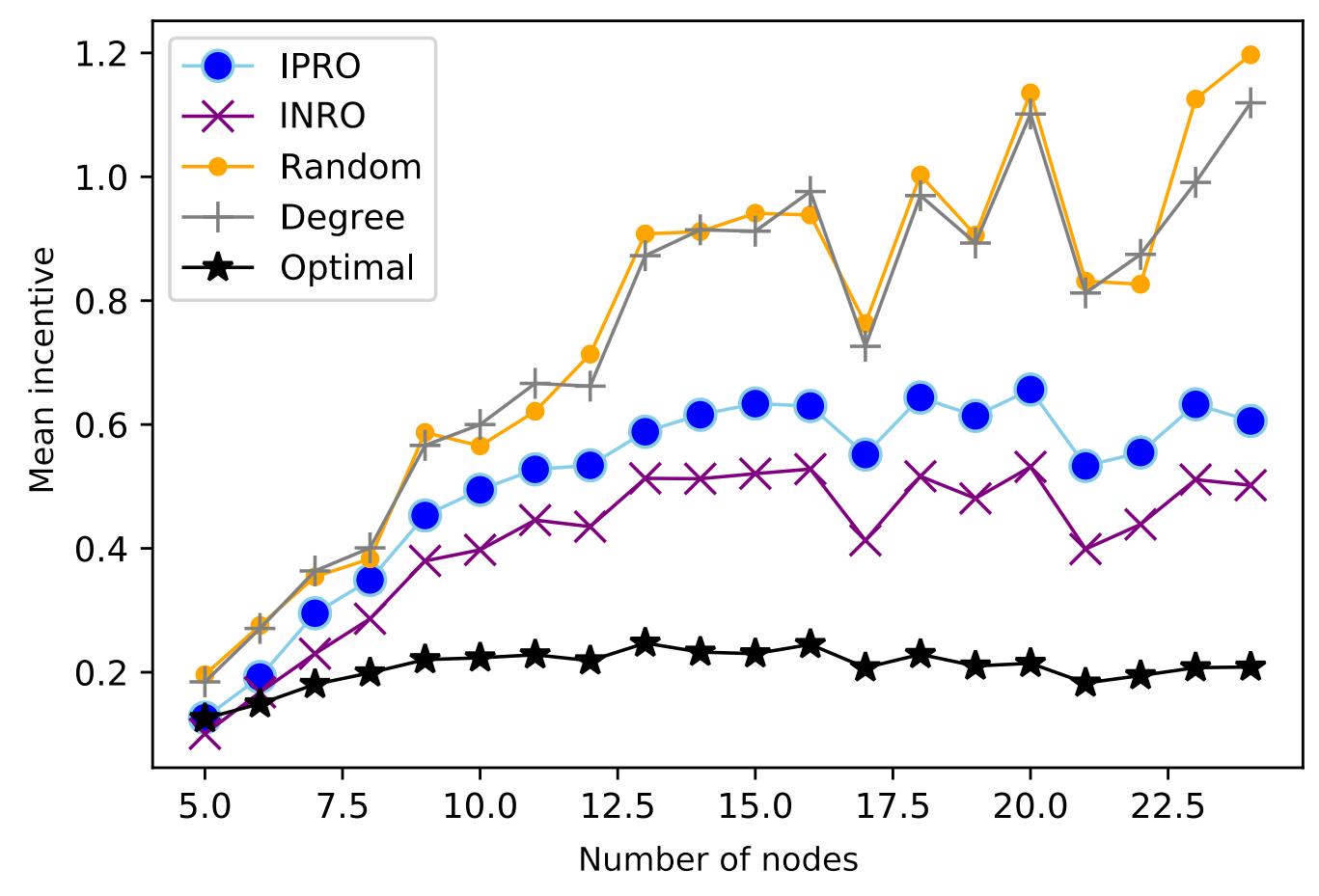}
    \caption{Performance comparison between five methods for Algorithm \ref{alg:targeted} for populations of different sizes on a network of opponent-coordinating imitators.}\label{fig:sim_im2}
    \end{subfigure}
    \\
    \begin{subfigure}[b]{0.45\textwidth}
            \includegraphics[width=\linewidth]{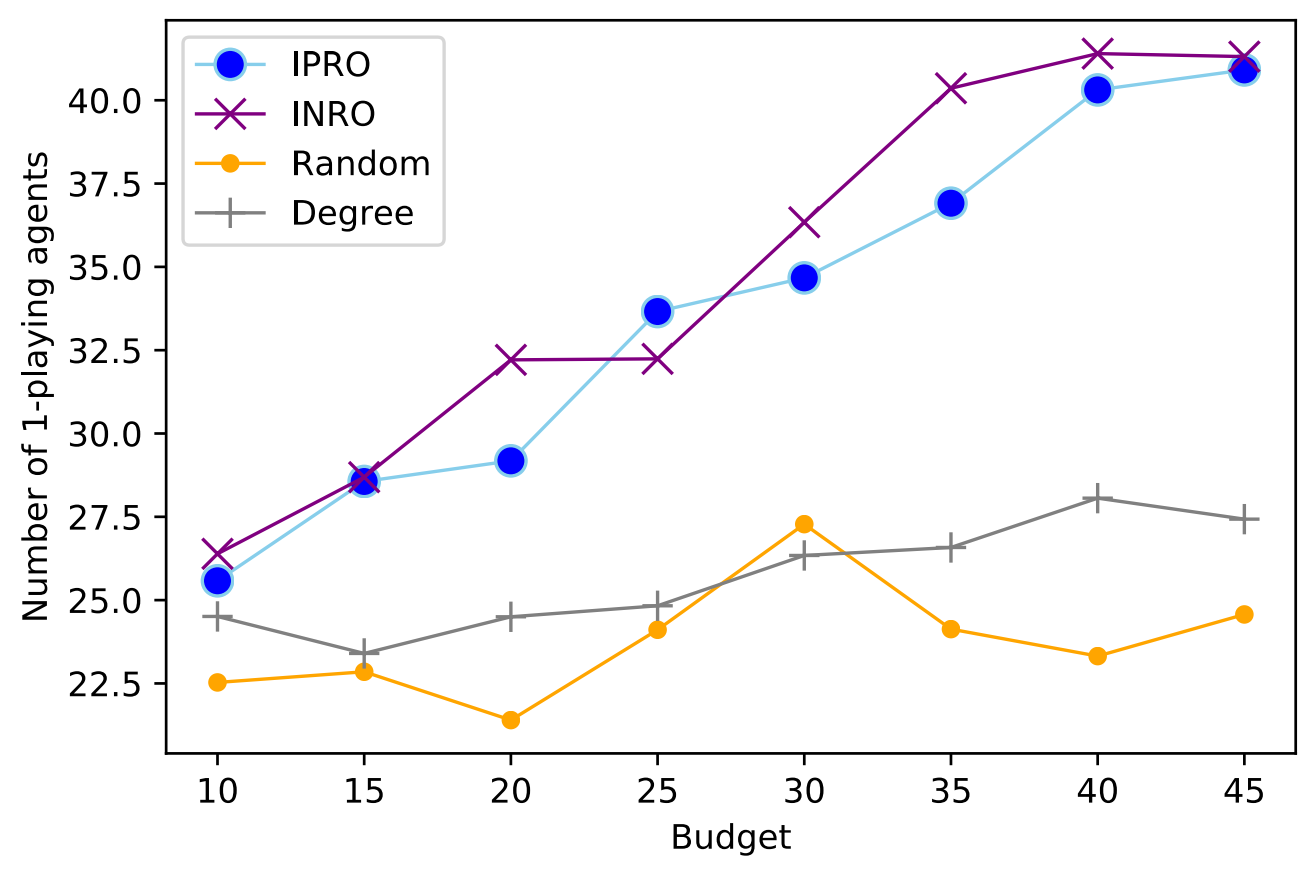}
    \caption{Performance comparison between four methods in solving Problem~\ref{prob:budgeted} for populations of size 100 on a network of opponent-coordinating imitators.}\label{fig:sim_im3}
    \end{subfigure}
    \caption{\textbf{Targeted reward control of networks of opponent-coordinating imitators.}}\label{fig:sim_im}
\end{figure}

\begin{figure}[ht]
\centering
    \begin{subfigure}[b]{0.45\textwidth}
            \includegraphics[width=\linewidth]{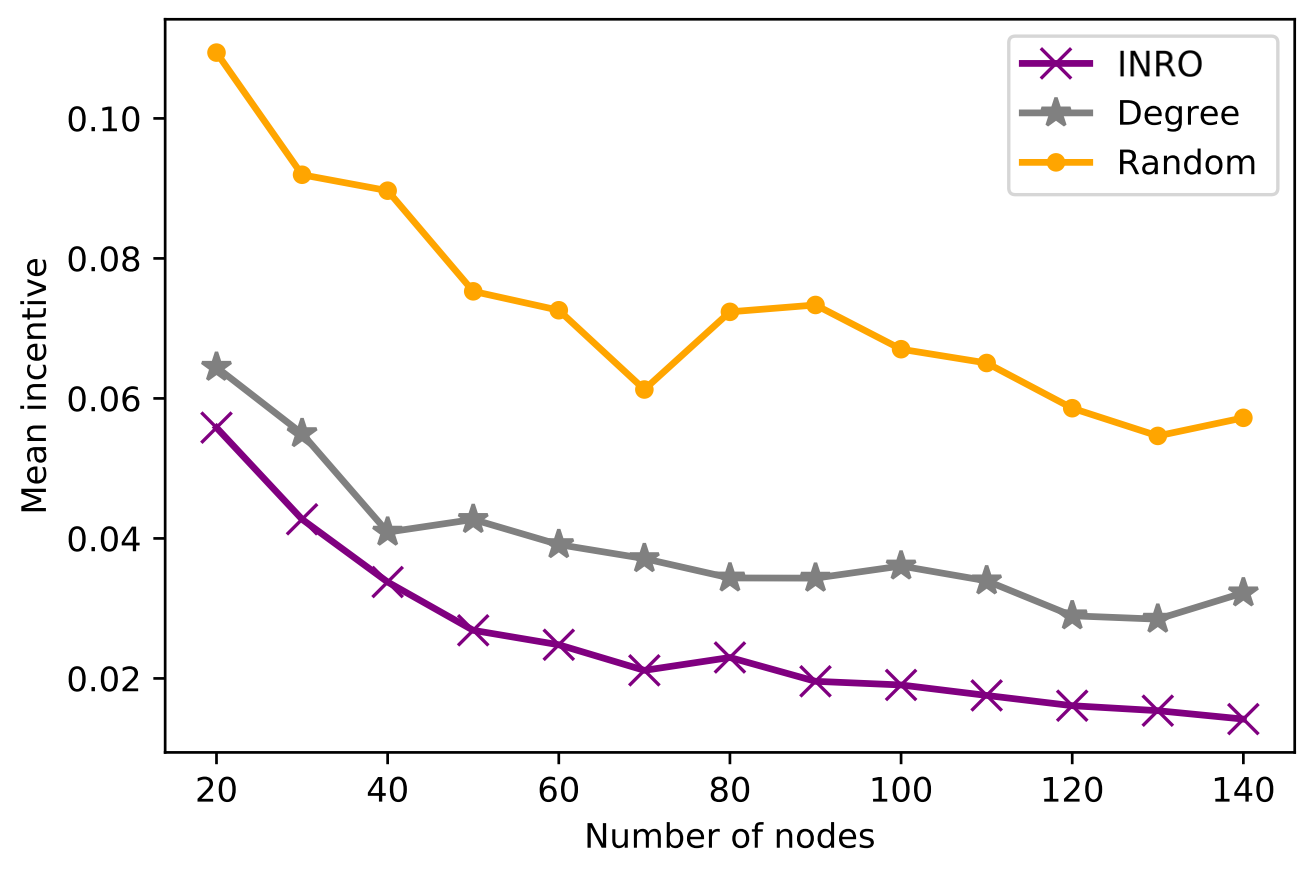}
    \caption{Performance comparison between three methods in Algorithm \ref{alg:targeted} for populations of different sizes on a network of coordinating imitation and best response agents.}\label{fig:sim_mix1}
    \end{subfigure}
    \\
    \vspace{15pt}
\begin{subfigure}[b]{0.45\textwidth}
            \includegraphics[width=\linewidth]{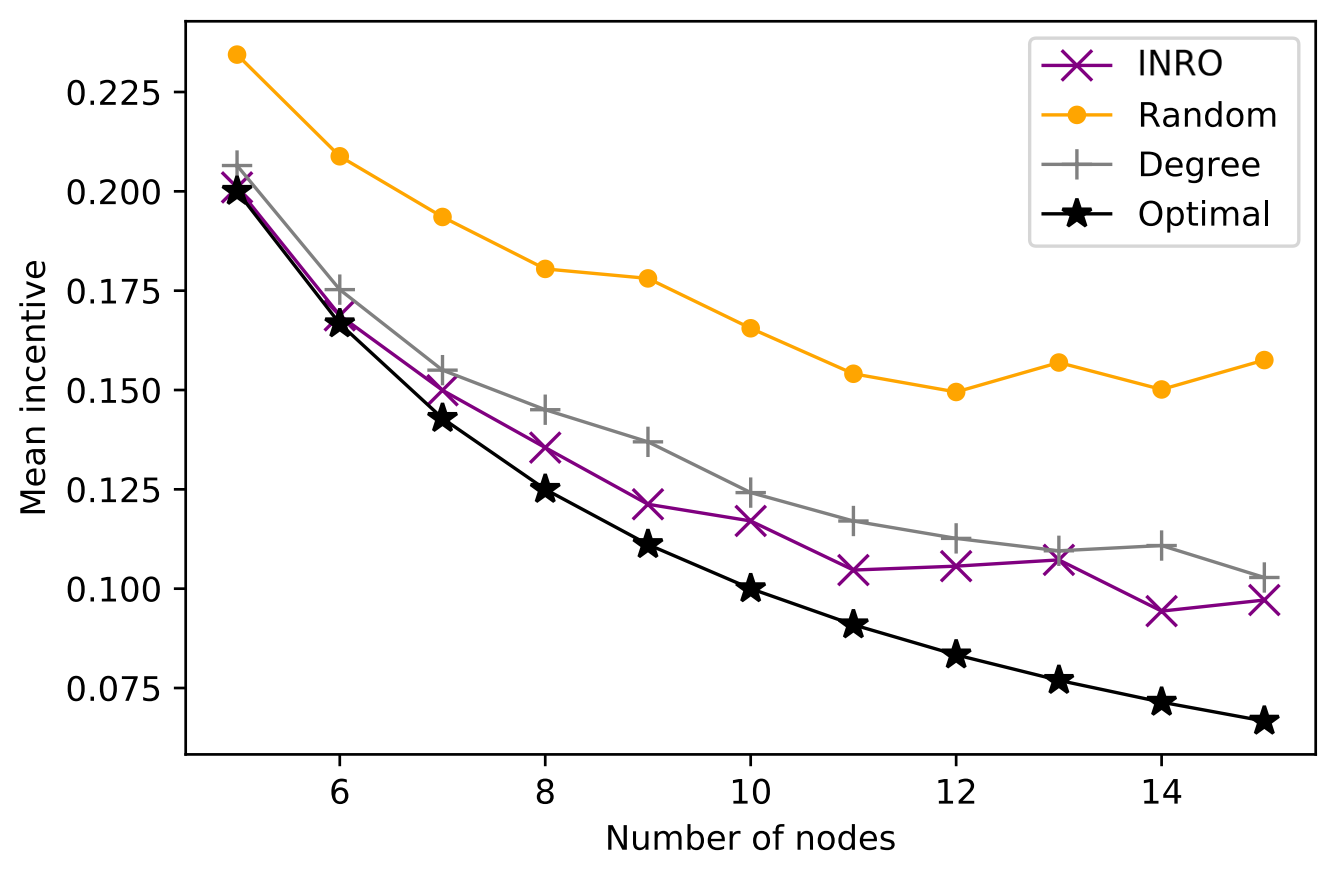}
    \caption{Performance comparison between four methods in Algorithm \ref{alg:targeted} for populations of different sizes on a network of coordinating imitation and best response agents.}\label{fig:sim_mix2}
    \end{subfigure}
    \\
    \begin{subfigure}[b]{0.45\textwidth}
            \includegraphics[width=\linewidth]{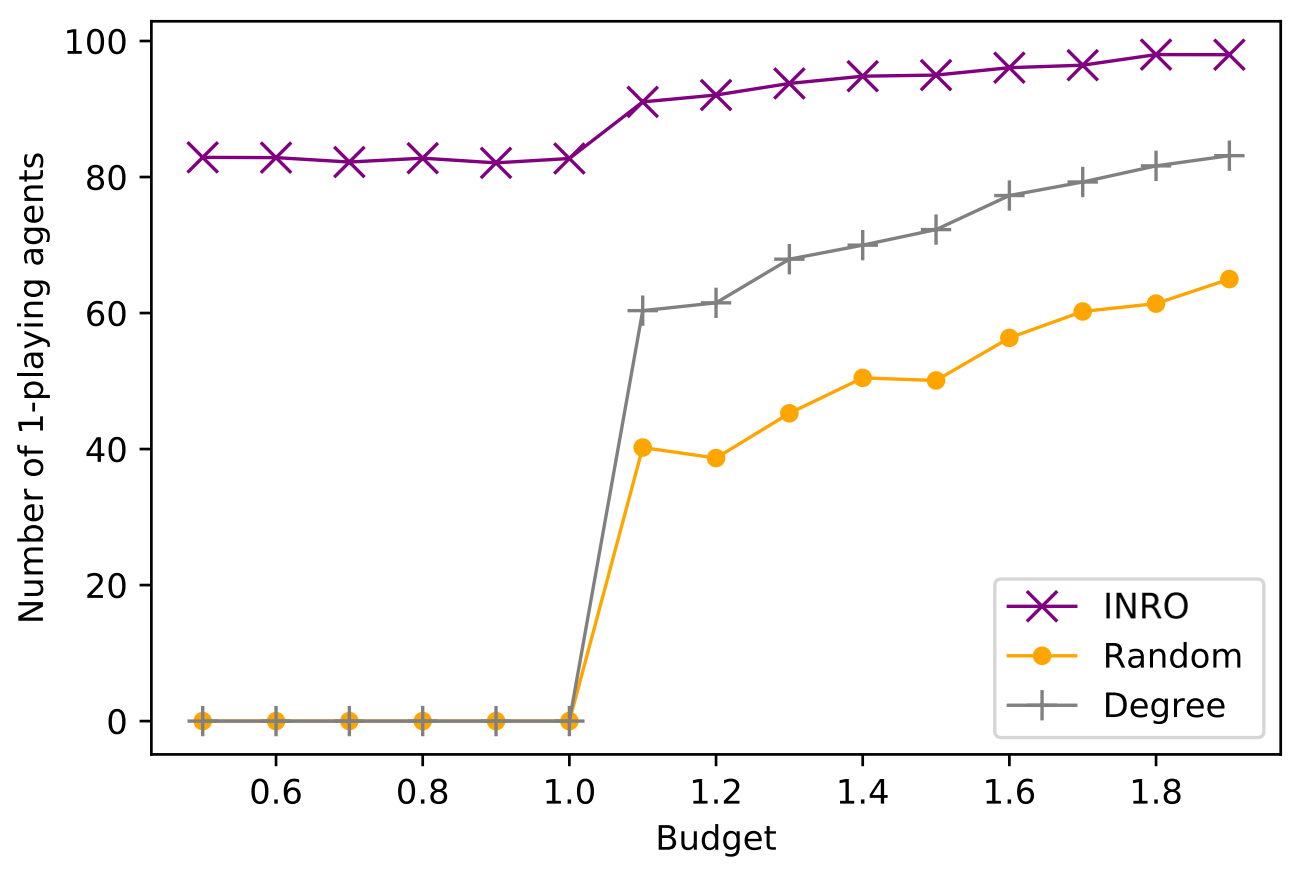}
    \caption{Performance comparison between three methods in solving Problem~\ref{prob:budgeted} for populations of size 100 on a network of coordinating best response and imitation agents.}\label{fig:sim_mix3}
    \end{subfigure}
    \caption{\textbf{Targeted reward control of mixed networks of best-responders and imitators.}}\label{fig:sim_mix}
\end{figure}

We present nine simulation results over Figures \ref{fig:sim_br} to \ref{fig:sim_mix} for studying the following three effects over the aforementioned three networks:
\begin{itemize}
    \item \textbf{(Problem \ref{prob:targeted}) The effect of population size on the amount of incentive needed.} 
    Tested over 100 networks of size 20 to 140 (Figures~\ref{fig:sim_br1}, \ref{fig:sim_im1} and \ref{fig:sim_mix1}).
    \item \textbf{(Problem \ref{prob:targeted}) The effect of population size on the amount of incentive needed.} 
    Tested over 100 networks of size 5 to 15 (Figures~\ref{fig:sim_br2}, \ref{fig:sim_im2} and \ref{fig:sim_mix2}).
    The maximum size was limited to 15 as finding the optimal solution is costly.
    \item \textbf{(Problem \ref{prob:budgeted}) The effect of budget on the number of agents playing the desired strategy.} 
    Tested over 100 networks of size 100 (Figures~\ref{fig:sim_br3}, \ref{fig:sim_im3} and \ref{fig:sim_mix3}).
\end{itemize}

The simulations were programmed in \texttt{Python}, and all codes are available at the \texttt{Git} repository \href{https://github.com/NegarSa/network-games}{\texttt{https://github.com/NegarSa/network-games}}.


\section{Conclusion}
We proposed the notion of coordinating revision protocols that dictates the active agent to switch to an option only if another agent has switched to that option. 
A population governed by coordinating revision protocols, i.e., a coordinating population, is guaranteed to equilibriate under a random activation sequence; however, it may never equilibrate under a non-random activation sequence. 
Therefore, coordinating populations do not admit a Lyapunov function in general, although they may admit ``Lyapunov-like'' functions that are defined over a particular subsequence of the activation sequence \cite{ramazi2017asynchronous}. 
This highlights the need for new tools, other than conventional Lyapunov-based methods, for studying decision-making dynamics.

Despite including a wide range of dynamics, especially network games, some populations such as best-responders associated with a positive diagonal payoff matrix, are not coordinating. 
The extension of coordinating populations to such cases remains an open problem. 

We showed that neither supermodular nor coordinating utilities are a special case of the other.
However, being defined on revision protocols rather than utilities, the notion of coordinating revision protocol covers a wider range of decision-making populations, such as opponent-coordinating imitation populations, that are not supermodular in general. 
Necessary and sufficient conditions for the two to be equivalent remains an open problem. 

Finally, simply based on the total number of agents choosing the desired strategy, we designed the new INRO algorithm that preforms competitively with previous algorithms for best-responders and imitators (IPRO), and applies to more general coordinating populations, such as the mixture of the two, as well as many yet not investigated decision-making dynamics. 

Our results on coordinating populations cover previously established convergence results for coordinating best-responders. 
We know that anticoordinating best-responders also converge; however, they are not coordinating. 
Is it possible to define a general notion opposite to coordinating populations that covers the anticoordinating best-responders case? 
This also remains concealed.

\bibliographystyle{IEEE}
\bibliography{bib}

\appendix
\begin{lemma} \label{proofOfExample:coo}
    The population in Example \ref{example:coo} reaches an equilibrium starting from any initial state and under any activation sequence.
\end{lemma}
\begin{IEEEproof}
    First, construct the nearest neighbor digraph of the points in the plane $\mathbb{D} = (\N, \E)$, $\E = \{(u, v) | v = \argmin_{k \in \N} d(u, k)\}$.
    In this directed graph, each weakly connected component has exactly one cycle, and its length is two \cite{nearest}.
    Consider one of the weakly connected components in the graph.
    The first time one of the agents in the 2-cycle is activated, say time $t_0$, the choices in that cycle will never change, and we can remove the out-going edges from both of the agents. 
    If there is an agent $i$ that has a directed edge to one of the agents in the 2-cycle, the first time $i$ is activated after $t_0$, her choice will never change again, and we can remove the out-going edge from $i$.
    Then we can repeatedly identify an agent who has zero out-degree but a positive in-degree, and activating the node that is linked to her, we make the nodes choices fixed and can remove their out-going edges.
    This will eventually empty the cluster, and the same can be done for other weakly connected components, resulting in an equilibrium. 
\end{IEEEproof}

\begin{IEEEbiographynophoto}{Negar Sakhaei}
received her B.Sc. in Computer Engineering from Isfahan University of Technology in 2020. She is currently pursuing a Master's degree from Sharif University of Technology. 
\end{IEEEbiographynophoto}

\begin{IEEEbiographynophoto}{Zeinab Maleki}
is currently an assistant professor at the Department of Electrical and Computer Engineering at Isfahan University of Technology, Iran. She received her MSc and PhD degrees, both in mathematics from Isfahan University of Technology in 2008 and 2015 respectively. She granted a general membership to spend the 2014-2015 academic year at the Institute for Mathematics and its Applications (IMA) at the University of Minnesota. From 2015 to 2016 she was a postdoctoral fellow in the Combinatorics and Computing Group at Institute for Research in Fundamental Science in Tehran. Since 2016 she has been with the Department of Electrical and Computer Engineering at Isfahan University of Technology as assistant professor.
\end{IEEEbiographynophoto}

\begin{IEEEbiographynophoto}{Pouria Ramazi}
	is currently an assistant professor at the Department of Mathematics and Statistics at Brock University, Canada. 
	He received the B.S. degree in electrical engineering in 2010 from University of Tehran, Iran, the M.S. degree in systems, control and robotics in 2012 from Royal Institute of Technology, Sweden, and the Ph.D. degree in systems and control in 2017 from the University of Groningen, The Netherlands. 
	He was a postdoctoral research associate with the Departments of Mathematical and Statistical Sciences and Computing Science at the University of Alberta from August 2017 to November 2020.
\end{IEEEbiographynophoto}


\end{document}